\documentclass[useAMS,usenatbib]{mn2e}
\usepackage{graphicx}
\title[Biomolecules in Sagittarius B2]
{A Search for Biomolecules in Sagittarius B2 (LMH) with the ATCA}
\author[P. A. Jones et al.]{P. A. Jones$^{1}$,
\thanks{E-mail: Paul.Jones@csiro.au (PAJ); Maria.Cunningham@unsw.edu.au (MRC)} 
M. R. Cunningham$^{1}$,
P. D. Godfrey$^{2}$ and D. M. Cragg$^{2}$ \\
$^{1}$School of Physics, University of New South Wales, NSW 2052, Australia \\
$^{2}$School of Chemistry, PO Box 23, Monash University, Clayton, Victoria 3800, 
Australia }

\begin{document}

\date{Accepted . Received ; in original form }

\pagerange{\pageref{firstpage}--\pageref{lastpage}} \pubyear{2006}

\maketitle

\label{firstpage}

\begin{abstract}

We have used the Australia Telescope Compact Array to conduct a search for the simplest amino acid, glycine (conformers I and II), and the simple chiral molecule propylene oxide at 3-mm in the Sgr B2 LMH. We searched 15 portions of spectrum between 85 and 91 GHz, each of 64 MHz bandwidth, and detected 58 emission features and 21 absorption features, giving a line density of 75 emission lines and 25 absorption lines per GHz stronger than the 5$\sigma$ level of 110 mJy. Of these, 19 are transitions previously detected in the interstellar medium, and we have made tentative assignments of a further 23 features to molecular transitions. However, as many of these involve molecules not previously detected in the ISM, these assignments cannot be regarded with confidence. Given the median line width of 6.5 km/s in Sgr B2 LMH, we find that the spectra have reached a level where there is line confusion, with about 1/5 of the band being covered with lines. Although we did not confidently detect either glycine or propylene oxide, we can set 3$\sigma$ upper limits for most transitions searched. We also show that if glycine is present in the Sgr B2 LMH at the level of $N = 4 \times 10^{14}$ cm$^{-2}$ found by Kuan et al. (2003) in their reported detection of glycine, it should have been easily detected with the ATCA synthesized beam size of 17.0 x 3.4 arcsec$^2$, if it were confined to the scale of the LMH continuum source ($< 5$ arcsec). This thus puts a strong upper limit on any small-scale glycine emission in Sgr B2, for both of conformers I and II.

\end{abstract}

\begin{keywords}
ISM:molecules - radio lines:ISM - ISM:individual:Sgr B2
\end{keywords}

\section{Introduction}

It is generally presumed that life on Earth developed from an initial reservoir of simpler 
prebiotic organic material such as amino acids, fatty acids and sugars. What is not clear is whether the chemical evolution to produce this initial reservoir occurred on the early Earth, in the pre-solar nebula, or even earlier in interstellar molecular clouds. In fact, a significant body of work exists which suggests that the necessary chemical evolution could not have taken place on the early Earth itself (Joyce et al. 1984; Bailey et al 1998), suggesting that the chemical evolution took place within  the pre-solar nebula, with the organic material being delivered to the Earth by meteorites and comets during the phase of bombardment.

Although there is expected to be some chemical and physical processing of material originating in the ISM, through the pre-solar nebula, into cometary and meteoritic material, and on to the surface of a planet, a study of the most complex molecules in the ISM is an important step in determining how likely it is that the reservoir of molecules from which life evolved originated in the pre-solar nebula. Indeed, a large number of surprisingly complex molecules, including amino acids, have been found in carbonaceous chondrite meteorites (see e.g. Wirick et al. 2006), which are thought to be relatively unprocessed remnants of the presolar nebula.

\begin{figure} 
\begin{picture}(80,100)(1,2)
\put(30,47) {CH$_2$}
\put(80,47) {CH}
\put(125,47) {CH$_3$}
\put(58,80) {O}
\put(39,59){\line(1,1){19}}
\put(54,51){\line(1,0){20}}
\put(101,51){\line(1,0){20}}
\put(85,59){\line(-1,1){19}}
\end{picture}
\caption{The chemical structure of the simple chiral molecule propylene oxide ($\rm{c-CH_{3}C_{2}H_{3}O}$)}
\label{fig1}
\end{figure}
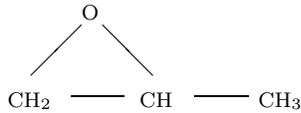

In this paper we
describe an Australia Telescope Compact Array (ATCA) search for two biologically important molecules:  the simplest amino acid glycine ($\rm{NH_{2}CH_{2}COOH}$), and the cyclic molecule propylene oxide ($\rm{c-CH_{3}C_{2}H_{3}O}$) which is chiral. Chiral molecules are those where asymmetry in carbon atom placement within a 
molecule leads to two distinct forms (mirror images or enantiomers). Figure \ref{fig1} shows the chemical structure of proylene oxide.

\subsection{Other Searches for Glycine}

There have been many searches for glycine in molecular clouds, dating back over
20 years (eg. Brown et al. 1979, Hollis et al. 1980, Snyder et al. 1983) with 
negative or ambiguous results. 
Miao et al. (1994) and Snyder (1997) reported a possible 
detection of a 90 GHz transition of glycine in Sgr B2 (N) with BIMA and OVRO
(interferometers). There were some caveats in Snyder (1997) due to uncertainty 
in the glycine line rest frequency, the different velocity components in Sgr 
B2 (N), and the confusion with other lines. Combes, 
Rieu \& Wlodarczak (1996) used the IRAM 30-m telescope (single dish) for 
the positions Orion SiO and Sgr B2 (OH) and concluded that the confusion 
of lines in the millimetre spectra, particularly in the Orion position, 
could make the detection of glycine impossible, as the glycine lines would
be below the confusion limit. 

Kuan et al.(2003, 2004a, 2004b) have reported the detection of 
glycine in Sgr B2 (N-LMH), Orion KL and W51 e1/e2 using the NRAO 12-m telescope.
They used the detailed comparison of lines in the different sources to help
solve the problem of ``interloper'' lines of other species. The use of rotation diagrams to provide a quantitative comparision of many detections and upper limits of the glycine lines  for all three sources gives greater confidence in the results. The total column density of glycine conformer I determined by Kuan et al. (2003) in Sgr B2 (N-LMH) is $N = 4 \times 10^{14}$ cm$^{-2}$ and rotational temperature $T = 75$~K. 

This reported detection of Kuan et al. (2003) has been disputed in a 
paper by Snyder et al.(2005), who
use unpublished NRAO 12-m data, and published SEST data from Nummelin et al.
(1998) to show that upper limits from non-detections of glycine lines quite
stringently rule out glycine in the three sources at the level reported by Kuan
et al. (2003) unless present as the higher energy conformer II. In addition, they argue how random coincidences of lines 
misidentified as glycine could lead to a plausible looking, but spurious,
fit to lines in a rotation diagram.

\subsection{The Importance of Chiral Molecules}

Investigations of primitive mechanisms of self-replication that involve RNA only, indicate the need for a reservoir of homochiral (having like chirality) 
organic molecules on the early Earth, before living organisms could evolve 
(Joyce et al. 1984). An excess of about 10 \% in one enantiomer over the other 
(L over D in the case of amino acids on the Earth) is needed for RNA template 
reproduction to work. 

Non-biological processes always create chiral organic molecules in a racemic 
mix (equal numbers of both handednesses) and there are no known abiotic 
mechanisms operating on the surface of the Earth to generate the required 
enantiomeric excess. Therefore, an extraterrestrial origin for the prebiotic 
molecules needed to seed life on Earth is likely if we accept the hypothesis 
that primitive reproduction involved only RNA. 

An extraterrestrial mechanism for producing the requisite homochirality is 
necessary. The most likely mechanism is circularly polarized UV radiation in 
the ISM, which is known to be capable of generating an enantiomeric excess by 
selectively destroying one enantiomer over the other (Bailey et al 1998).

Regrettably radio astronomy cannot distinguish the mirror image molecular forms,
which in different concentrations would signal chiral imbalance, but it can 
potentially detect the presence of molecules capable of this property. To date, no molecules possessing a chiral centre have been detected in the ISM. Glycine is not a chiral molecule, and our search for biologically important chiral molecules with the ATCA and Mopra is focusing on propylene oxide.  The simpler, non-chiral, but chemically related cyclic species ethylene oxide ($\rm{c-C_{2}H_{4}O}$) has already been detected in Sgr B2N (Dickens et al. 1997).

\subsection{Our Search Strategy}

Two developments prompted this search for glycine and propylene oxide:

\begin{itemize}
 
\item  Firstly, we have been able to use laboratory measured values of the rest frequencies of glycine and propylene oxide transitions between 85 and 115 GHz rather than the less accurate calculated  values (Lovas et al. 1995) on which previous searched at 3 mm have had to rely. The measurements were made at the School of Chemistry, Monash University, using a purpose-built Stark-modulated free-jet microwave absorption spectrometer of a design similar to that described previously (Brown et al. 1988).  These measured values have an uncertainty of around only 0.07 km/s in the central velocity of a transition \footnote{The new laboratory measurements are not yet publically available. Enquiries can be addressed to the authors}.

\item Secondly, the new capabilities of the ATCA at millimetre wavelengths allows a search at high spatial resolution, minimising beam dilution effects and, as an interferometer, filtering out confusing lines from the extended gas within the beam. 

\end{itemize}

Our search has a multi-pronged approach, involving both single dish and 
interferometer observations of multiple frequencies within the 3-mm band.
The position Sgr B2 (LMH) is of small diameter ($<$5 arcseconds) and has so 
far provided detections of the related complex molecules methyl formate and 
acetic acid (which is a precursor to glycine in some reaction networks). 
These molecules seem to be confined to the LMH. However, another similar 
molecule, glycolaldehyde (Hollis et al. 2001), was found to be extended in the Sgr B2 (N) 
molecular cloud, with very low concentrations in the LMH. These three 
molecules are believed to have a common origin with glycine (Sorrell 2001) 
based on a model of chemical reactions in icy grain mantles. It is therefore 
important to search for glycine in both extended and compact dense 
molecular gas.

We are thus using both the ATCA and the single dish Mopra radio telescope. The 
Mopra telescope, jointly operated by the ATNF CSIRO, and the 
University of New South Wales, Australia, has a 22-m diameter, making it the 
largest single dish telescope operating at 3-mm in the southern hemisphere. 
Mopra results will be discussed in a subsequent paper.

In the laboratory, glycine can adopt more than one conformation or spatial
arrangement of the groups. Since it is uncertain which conformer of glycine
may be present in the ISM, we searched for lines from both conformer I and 
conformer II (Lovas et al. 1995). 
Conformer I has lower energy than II by around $E/hc = 700$~cm$^{-1}$
(Lovas et al. 1995), but conformer II has a larger dipole moment and hence
stronger lines. 
There is a significant  energy barrier between conformers 
I and II, since to go from one to the other requires breaking a hydrogen bond
rather than just an internal rotation. The energy barrier is estimated
theoretically at around $E/hc = 5000$~cm$^{-1}$ or $E/k = 7000 K$ 
(Godfrey, Brown \& Rodgers 1996). At interstellar temperatures the rate of 
tunnelling through the barrier will be low. If glycine is formed at low 
temperatures on grain surfaces, it may be stuck in the original conformation,
and if the formation chemical processes favoured conformer II, it would
relax only very slowly into the lower energy conformer I.
Thus it is worthwhile searching for both conformers.

\section[]{ATCA Observations and Data Reduction}

Observations were made with the Australia Telescope Compact Array, in 2002
and 2003, during the southern Winter period of conditions appropriate for 3-mm
observations. The log of observations is given in table 1. 


\begin{table}
\caption{Log of ATCA observations, frequency bands and lines searched. The frequencies quoted are from new laboratory measurements made at the School of Chemistry, Monash University.  The corresponding transitions are listed in table 4.
The asterisk in column 4 denotes the line used as rest frequency for the 
velocity scale when more than one line was in the observed band. ID specifies the
molecule, propylene oxide (PO) or glycine (G), and conformer, I or II (glycine only), with this transition frequency. Strength
is a measure of the intensity predicted at 10 K and 50 K (10 K if only
one number is given), relative to a value of 100 for the strongest
line of that species in the 80-115 GHz range. }
\begin{tabular}{ccclll}
\hline
Band  & Dates       & Arrays & Line Freq.  & ID &  Strength \\
Label &             & used   & (GHz) &    &  10/50 K \\
\hline
85477 & 2003 Jun 06 & H75   & 85.48422  & PO & 27  \\
      & 2003 Jul 03 & EW214 &           &                 &     \\
      &             &       &           &                 &     \\ 
86267 & 2003 Jun 06 & H75   & 86.262231* & G II & 27/51 \\
      & 2003 Jul 03 & EW214 & 86.283667 & G II & 27/51 \\
      &             &       &           &            &       \\
86374 & 2003 Jun 06 & H75   & 86.379992 & G II & 74/82 \\
      & 2003 Jul 03 & EW214 &           &            &       \\
      &             &       &           &            &       \\
86709 & 2002 Jun 06 & EW352 & 86.72160*  & PO & 29  \\
      & 2002 Aug 18 & H75   & 86.716051 & G II  &  74/82     \\
      &             &       &           &            &       \\
86754 & 2002 Jun 06 & EW352 &           &            &       \\
      &             &       &           &            &       \\
86880 & 2003 Jun 06 & H75   & 86.88593  & G I  & 80/61 \\
      & 2003 Jul 03 & EW214 &           &            &       \\
      &             &       &           &            &       \\
86954 & 2002 Jun 06 & EW352 & 86.960812 & G II & 74/89 \\
      & 2002 Aug 18 & H75   & 86.967057* & G II & 74/89 \\
      &             &       & 86.978582 & G II & 43/61 \\
      &             &       &           &            &       \\
88587 & 2003 Jun 07 & H75   & 88.59909  & PO &  8  \\
      & 2003 Jul 02 & EW214 & 88.60199* & PO &  8  \\
      &             &       &           &            &       \\
88761 & 2002 Aug 19 & H75   & 88.77871*  & PO & 15  \\
      &             &       & 88.792618 & G II &  4  \\
      &             &       &           &            &       \\
89502 & 2003 Jun 07 & H75   & 89.50043  & G I  & 62/67 \\
      & 2003 Jul 02 & EW214 & 89.51220*  & G I  & 22/44 \\
      &             &       & 89.53550  & G I  & 23/44 \\
      &             &       &           &            &       \\
89737 & 2003 Jun 07 & H75   & 89.75055  & G I  & 49/56 \\
      & 2003 Jul 02 & EW214 &           &            &       \\
      &             &       &           &            &       \\
89829 & 2002 Aug 19 & H75   & 89.87218  & PO  & 23 \\
      &             &       & 89.83184*  & G I  & 63/67 \\
      &             &       & 89.87571  & G I  & 34/52 \\
      &             &       &           &            &       \\
90022 & 2002 Aug 19 & H75   & 90.03589  & G I  & 25/30 \\
      &             &       & 90.04311*  & G I  & 63/72 \\
      &             &       & 90.04967  & G I  & 63/72 \\
      &             &       & 90.05689  & G I  & 25/30 \\
      &             &       &           &            &       \\
90451 & 2002 Aug 19 & H75   & 90.47516  & PO & 28  \\
      &             &       &           &            &       \\
90771 & 2003 Jun 07 & H75   & 90.78355  & G I  & 73/67 \\
      & 2003 Jul 02 & EW214 &           &            &       \\ 
\hline      
\end{tabular}
\label{tab1}
\end{table}

During this 2002 and 2003 period, the 3-mm system installed on the ATCA was 
an interim system using only three of the six ATCA antennas, rather than the 
five
antennas (on the 3-km EW track and NS spur) of the final system. Also, the
interim local oscillator (LO) system for the 3-mm band did not allow the full 
tuning range 85 - 105 GHz of the final system. 
In the interim system the LO was fixed
at 80.505 GHz, and the intermediate frequency (IF) passed through the
existing C-band (4.4 - 6.8 GHz) or X-band (8.0 - 10.8 GHz) filters, giving the 
sky frequency coverage 84.906 - 87.305 and 88.506 - 91.305 GHz. 

We selected the strongest predicted lines of glycine, conformers I and II, and propylene oxide in the available ATCA 3-mm frequency range, based on LTE calculations at 10 K and 50 K. The correlator configuration was two simultaneous bands of 64 MHz, each of 128 channels of bandwidth 0.5 MHz. Some of the bands included two or more biomolecule lines, so we covered 27 lines in 15 bands as shown in table 1.

Sagittarius B2, at declination -28$\degr$ passes nearly overhead at the ATCA, but
we did not observe below an elevation of around 30$\degr$, since the phase errors
and attenuation are worse at low elevation. The hour-angle coverage was therefore around 8
hours per observation, but the integration time was more like half this when 
the observing
overheads of pointing, phase calibrator, primary calibrator, bandpass calibrator 
and system temperature scans is taken
into account. We used B1730-130 as the phase calibrator, with a cycle time of
typically 10 minutes on source Sgr B2 and 2 minutes on B1730-130. Uranus was
used as the primary calibrator. Either 3C279 or 1921-293 were used as bandpass 
calibrators. We typically combined 2 days of observations
for each band, one with east-west spacings (EW214 or EW352) and one with
north-south spacings (H75), to get better \textit{u,v}--coverage. 
The range of antenna spacings was 30 to 120 m.

To increase the number of bands searched, albeit at the cost of reduced
sensitivity, we generally timeshared between two frequency
settings, getting four frequencies per observation (two settings times two
simultaneous bands).


\begin{table}
\caption{Resolution and sensitivity of the ATCA observations}
\begin{tabular}{lrrc}
\hline
Band  & Beamsize     &  PA      &  RMS      \\
Label & (arcsec$^2$) & ($\deg$) & (mJy/beam) \\      
\hline                                         
85477 &  7.0$\times$4.7 &  0  &  18 \\   
86267 &  7.0$\times$4.7 &  0  &  16 \\
86374 &  7.0$\times$4.7 &  0  &  16 \\  
86709 &  6.0$\times$3.5 &  0  &  14 \\ 
86754 & 13.5$\times$2.7 & -22 &  20 \\  
86880 &  7.0$\times$4.7 &  0  &  21 \\ 
86954 &  6.0$\times$3.5 &  0  &  15 \\  
88587 &  6.5$\times$5.0 &  0  &  35 \\   
88761 & 17.0$\times$3.4 & -84 &  33 \\  
89502 &  6.5$\times$5.0 &  0  &  26 \\   
89737 &  6.5$\times$5.0 &  0  &  26 \\   
89829 & 17.0$\times$3.4 & -81 &  26 \\  
90022 & 17.0$\times$3.4 & -83 &  22 \\  
90451 & 17.0$\times$3.4 & -85 &  31 \\  
90771 &  6.5$\times$5.0 &   0 &  27 \\   
\hline
\end{tabular}
\label{tab2}
\end{table}

The data were reduced with the MIRIAD package, ATNF version \\
(http://www.atnf.csiro.au/computing/software/miriad/). \\
The gain-elevation dependence was corrected with task {\it elevcor}, and 
the primary flux scale set using the data for Uranus and the task {\it plboot}.
There were some problems with the antenna baseline solutions for ATCA 3-mm
observations, particularly during 2002, so we used phase self-calibration 
using the strong continuum flux of Sgr B2 (LMH).

The continuum emission was subtracted from the {\it u,v} - data with the Miriad
task {\it imlin} by selecting the line-free channels. Continuum-free data cubes
were made for each of the 15 bands, and the data cubes cleaned. The size of the
clean restored beams are given in table 2. For the bands that had both EW and NS
ATCA arrays, the beams are roughly circular, elongated by a factor around 1.5
north-south, but for the five bands with only one ATCA array, the beams are
highly elongated north-south or east-west due to the rather poor 
{\it u,v} - coverage. 

The data cubes were generated with velocity coordinates, with the biomolecule 
line rest frequencies
given in table 1. Where two biomolecule lines are in the same band, the rest 
frequency used to convert from frequency to velocity is marked with an asterisk.

Continuum images were also made, using the line-free
channels, by combining data for all the bands observed in 2002 and in 2003.



\section{Results}
\subsection{Sgr B2 3-mm continuum}

The 3-mm continuum emission from Sgr B2 is resolved into two components by the 
ATCA, labelled N and N' by Kuan, Mehringer \& Snyder (1996) from 3-mm BIMA 
observations of similar resolution. With only three baselines in the ATCA data, 
and uncertain calibration, the continuum image is in fair agreement with 
previous continuum observations. For the combined continuum data for 2003, the 
total 
flux from Gaussian fits to the two components is 7.1 Jy, compared to a total of
6.9 Jy at 87.9 GHz in Kuan et al. (1996). However, the fluxes of the two
components in the 2003 ATCA image is 4.7 Jy for N and 2.4 Jy for N' compared to
3.7 and 3.2 Jy respectively at 87.9 GHz in Kuan et al. (1996), although Kuan et 
al. do find that N has a very steep spectrum (spectral index 
$\alpha = 4.6 \pm 0.5$,
where $S_{\nu} \propto \nu^{\alpha}$ so that the flux ratio is steeply 
frequency dependent). The positions from the ATCA 2003 continuum data are (J2000)
17 47 20.07, -28 22 18.2 for N and 17 47 20.39, -28 22 13.4 for N'. Due to the 
self-calibration applied, the absolute positions may be somewhat uncertain,
but the relative offset of N', 4.3 arcsec east and 4.7 arcsec north of N, 
is more reliable.

\subsection{Sgr B2 line emission and absorption}

The ATCA data cubes for Sgr B2 showed many line emission and absorption 
features. These features were spatially located close to the continuum position
N, so for display in figure 2 we show the spectra at the pixel close to the N
peak (0.5 arcsec west and 0.2 arcsec north of the fitted N continuum position
above). 

The RMS noise ($\sigma$) levels in the data cubes, obtained after clipping 
pixels at the $3 \sigma$ level to reduce bias, are given in table 2.
The values are in the range 14 to 35 mJy/beam, which is in good agreement with 
the expected thermal noise given the ATCA instrumental parameters (User's Guide 
to 3-mm Observing with the ATCA, 
http://www.atnf.csiro.au/observers/docs/3mm/index.html) and the integration 
time for the data sets.
Spectral-line features in the data cubes were considered significant at the
$5 \sigma$ level. Note that there are strong emission features (up to 8 Jy/beam 
in band labelled 89502), but also strong absorption features, such as that at -43 km/s in the band 
labelled 86754, which is identified with H$^{13}$CO$^{+}$. The continuum 
brightness temperature of component N is around 60 K, using the 
integrated flux 4.7 Jy, and deconvolved HPBW of 4.3 arcsec $\times$ 2.9
arcsec from the 2003 data, although somewhat uncertain due to the uncertainty in the 
deconvolved source size. As line absorption can occur if the line 
excitation temperature is less than this background brightness temperature, we consider both line absorption and emission features at the 
continuum peaks.

\begin{figure*}
\setlength{\unitlength}{1cm}
\begin{picture}(16,24)
\put(-1,23.5){\textbf{Figure 2.} Spectra of Sgr B2 (LMH) at the N continuum peak.}
\put(-1,23.1){The line identifications are as listed in Table 3.}
\put(7.8,24.4){\includegraphics{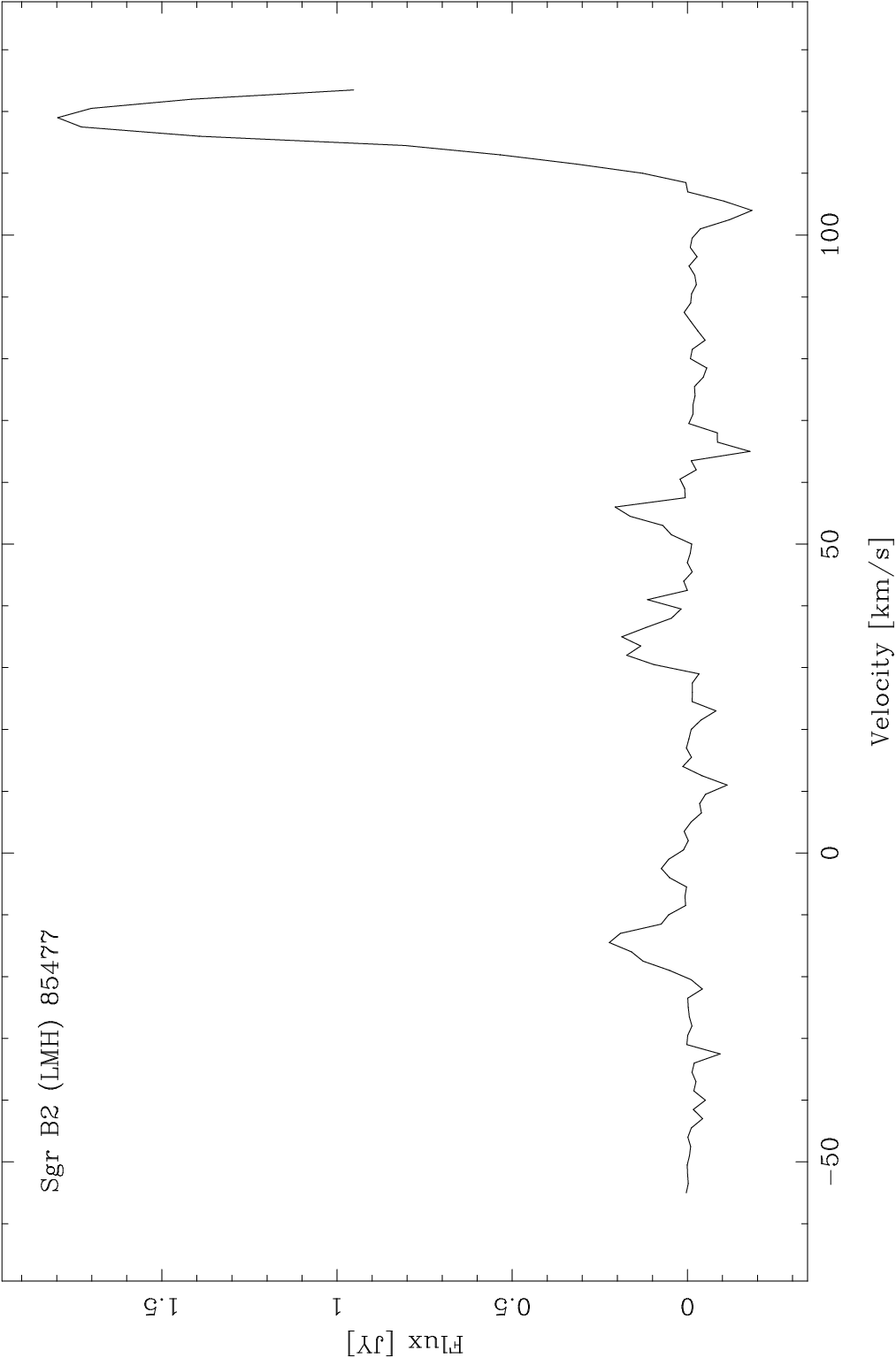}}
\put(15.75,23.455){\vector(0,-1){0.26}}
\put(14.25,23.2){\small ? C$_2$H$_3$CN}
\put(13.35,20.25){\vector(0,-1){0.4}}
\put(13.2,20.3){\small ? O$^{79}$BrO}
\put(11.655,19.75){\vector(0,-1){0.4}}
\put(11,19.9){\small ? HOONO$_2$}
\put(10.7,19.1){\vector(0,1){0.4}}
\put(10,18.76){\small U 85.506}
\put(-1,18.4){\includegraphics{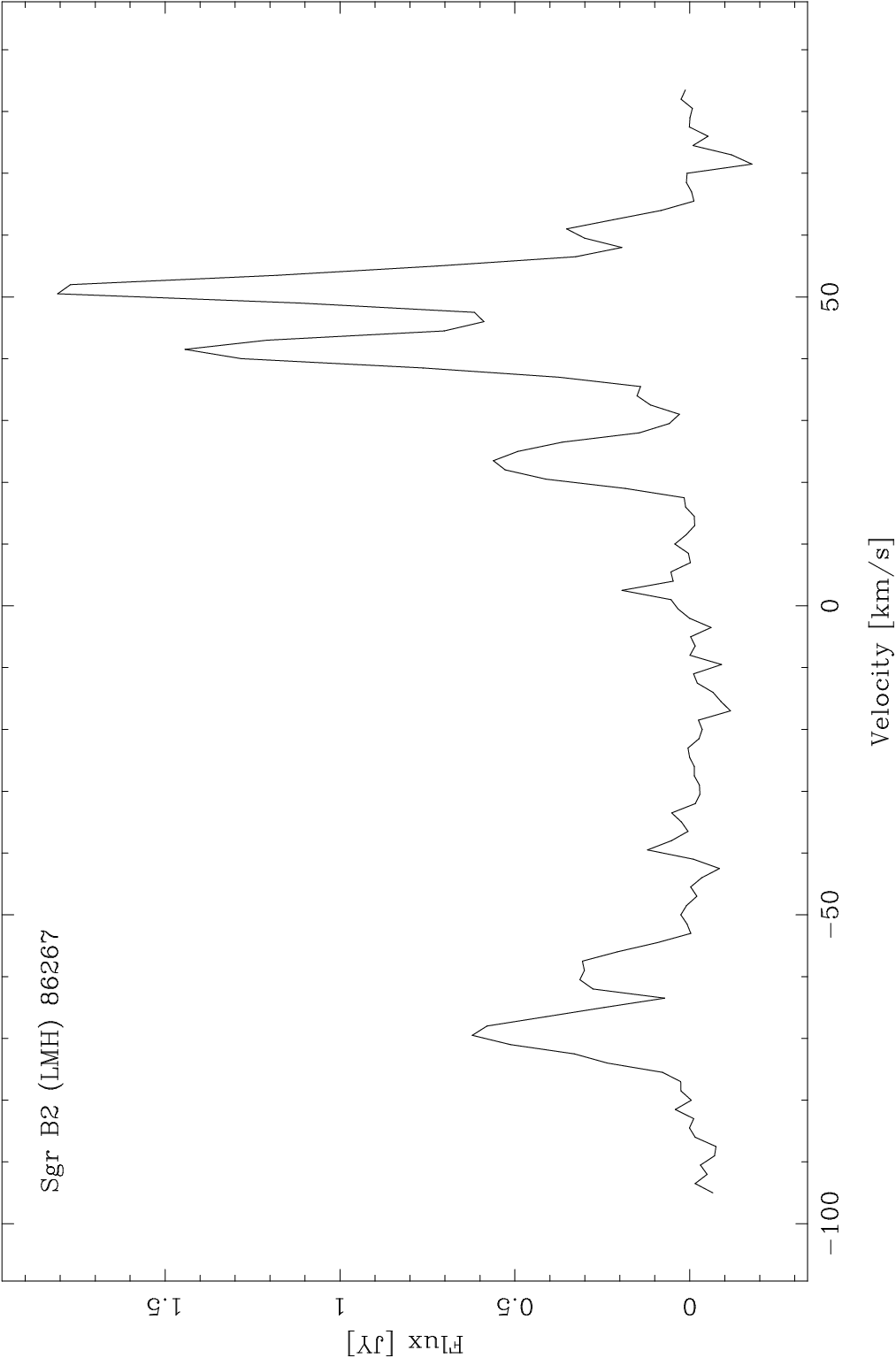}}
\put(5.89,17.65){\vector(0,-1){0.4}}
\put(5.1,17.735){\small CH$_3$OCHO}
\put(5.53,16.9){\vector(0,-1){0.4}}
\put(4,17){\small ? CH$_3$OCHO}
\put(4.85,14.98){\vector(0,-1){0.4}}
\put(3.75,15.1){\small C$_2$H$_5$OOCH}
\put(3.37,12.7){\vector(0,1){0.3}}
\put(3.45,12.76){\small NH$_2$CH$_2$CH$_2$OH}
\put(2.49,12.8){\vector(0,1){0.4}}
\put(1.4,12.8){\small O$^{37}$ClO}
\put(1.75,14.45){\vector(0,-1){0.4}}
\put(1.74,14.5){\small ? HOONO$_2$}
\put(1.36,15.16){\vector(0,-1){0.4}}
\put(0.9,15.22){\small ? HOONO$_2$}
\put(7.8,18.4){\includegraphics{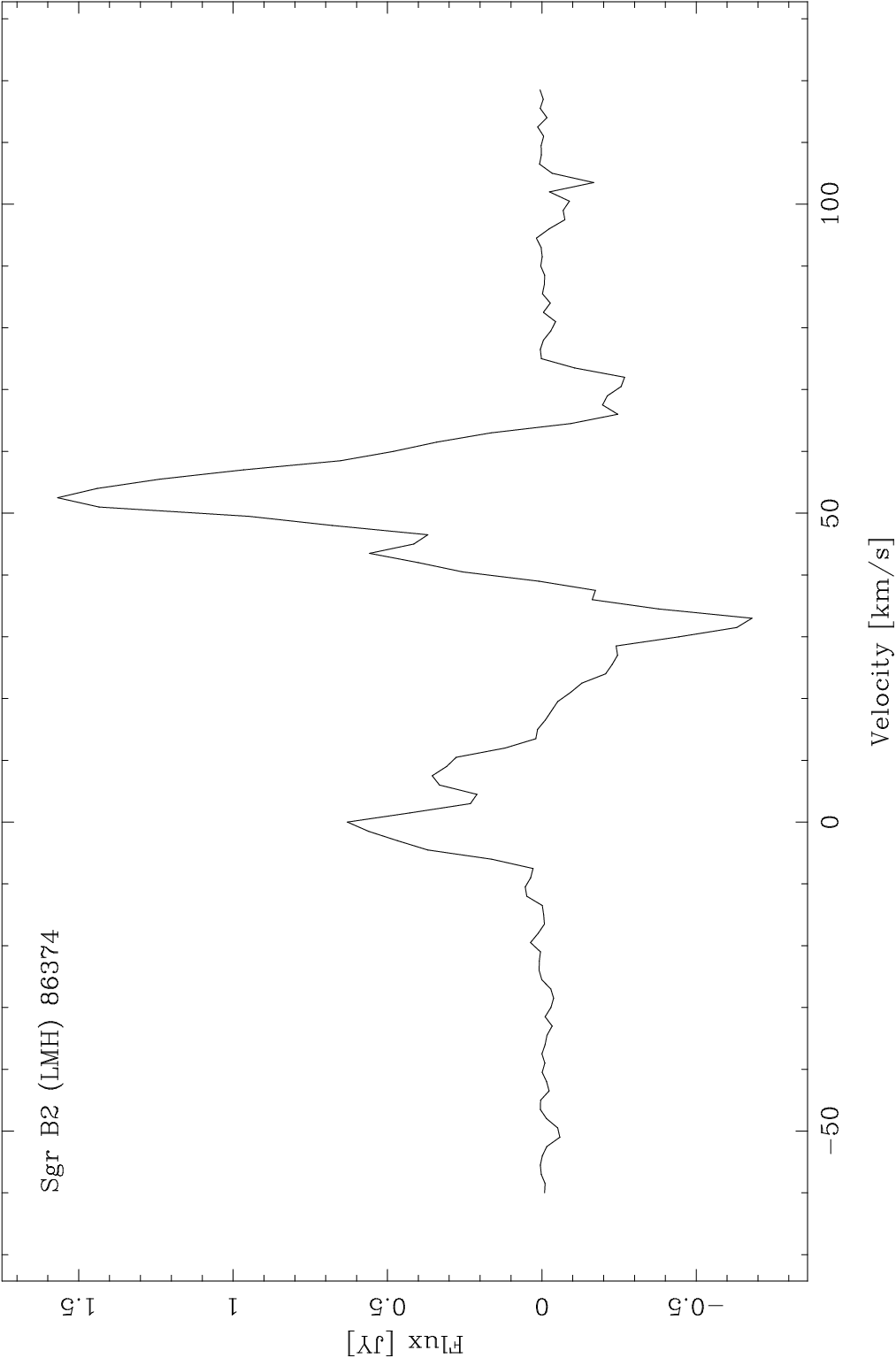}}
\put(13.42,17.48){\vector(0,-1){0.26}}
\put(13.5,17.25){\small ? N$_2$CHO}
\put(12.65,12.63){\vector(0,1){0.26}}
\put(11.35,12.75){\small ? CH$_2$F$_2$}
\put(-1,12.4){\includegraphics{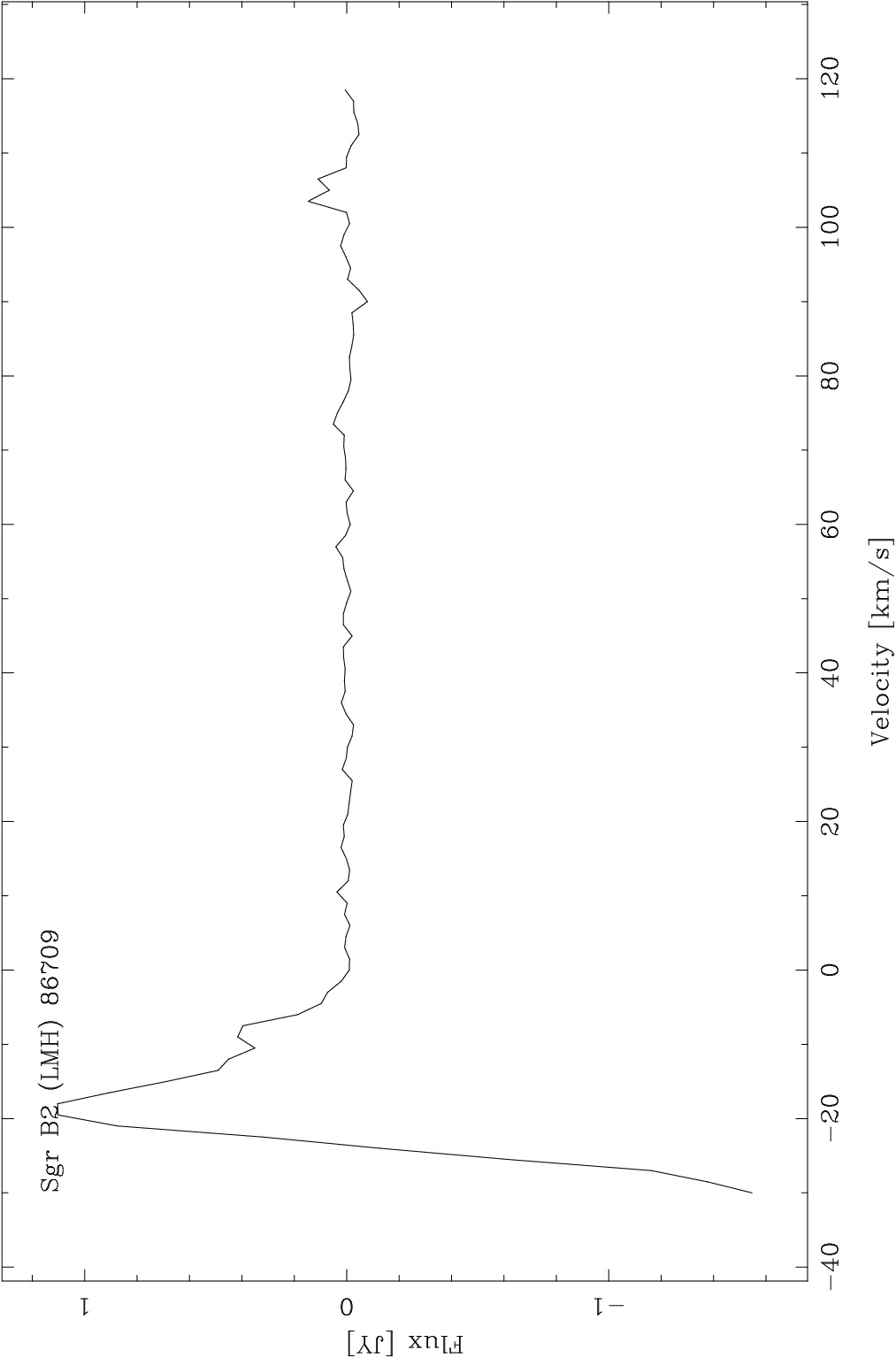}}
\put(6.45,10){\vector(0,-1){0.3}}
\put(6,10.1){\small ? O$^{79}$BrO}
\put(1.33,9.6){\vector(0,1){0.3}}
\put(1.6,9.72){\small CH$_3$CH$_2$CN}
\put(0.9,10.63){\vector(0,1){0.3}}
\put(1.1,10.75){\small CH$_3$CH$_2$CN}
\put(7.8,12.4){\includegraphics{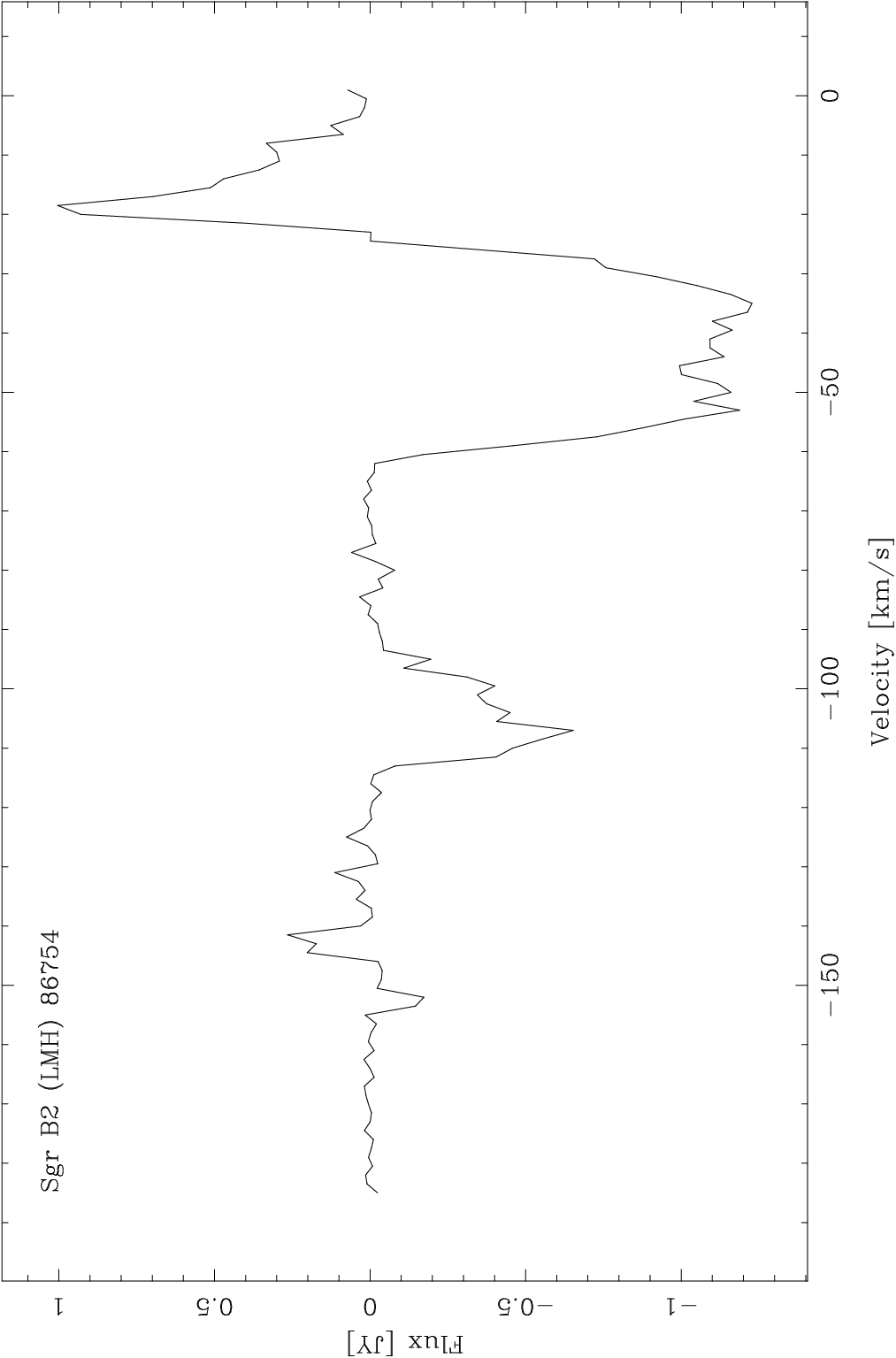}}
\put(15.25,9.95){\vector(0,1){0.3}}
\put(13.4,10){\small CH$_3$CH$_2$CN}
\put(15.55,9.3){\vector(0,1){0.4}}
\put(14.758,9){\small CH$_3$CH$_2$CN}
\put(14.3,6.63){\vector(0,1){0.3}}
\put(13,6.7){\small H$^{13}$CO$^+$}
\put(-1,6.4){\includegraphics{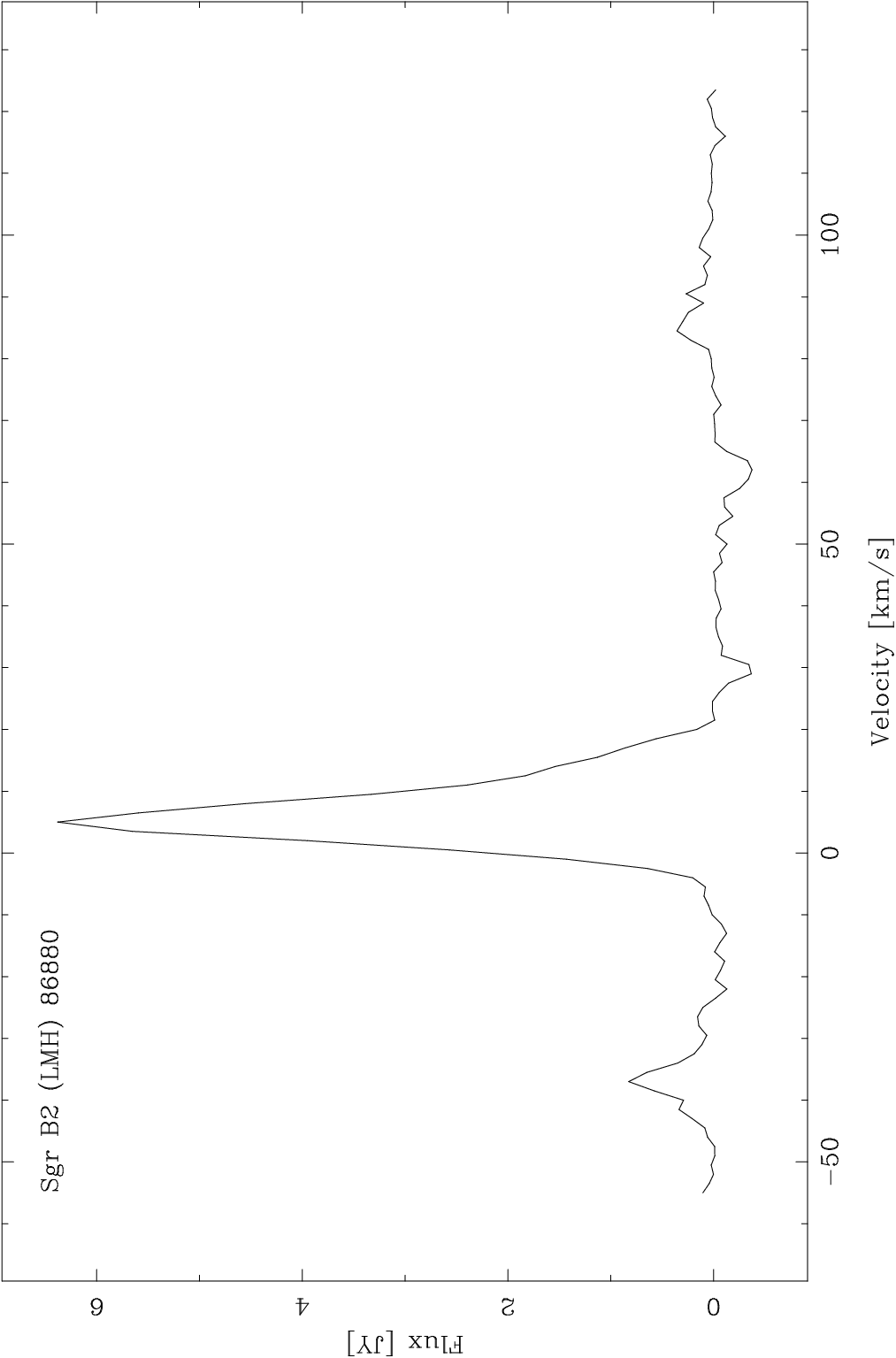}}
\put(5.65,0.9){\vector(0,1){0.3}}
\put(5.5,0.7){\small ? O$^{79}$BrO}
\put(3.55,1.5){\vector(0,-1){0.4}}
\put(3.5,1.6){\small ? HCOOD}
\put(2.85,1.9){\vector(0,1){0.3}}
\put(1.2,1.95){\small CH$_3$OH}
\put(2.55, 1.9){\vector(0,1){0.3}}
\put(7.8,6.4){\includegraphics{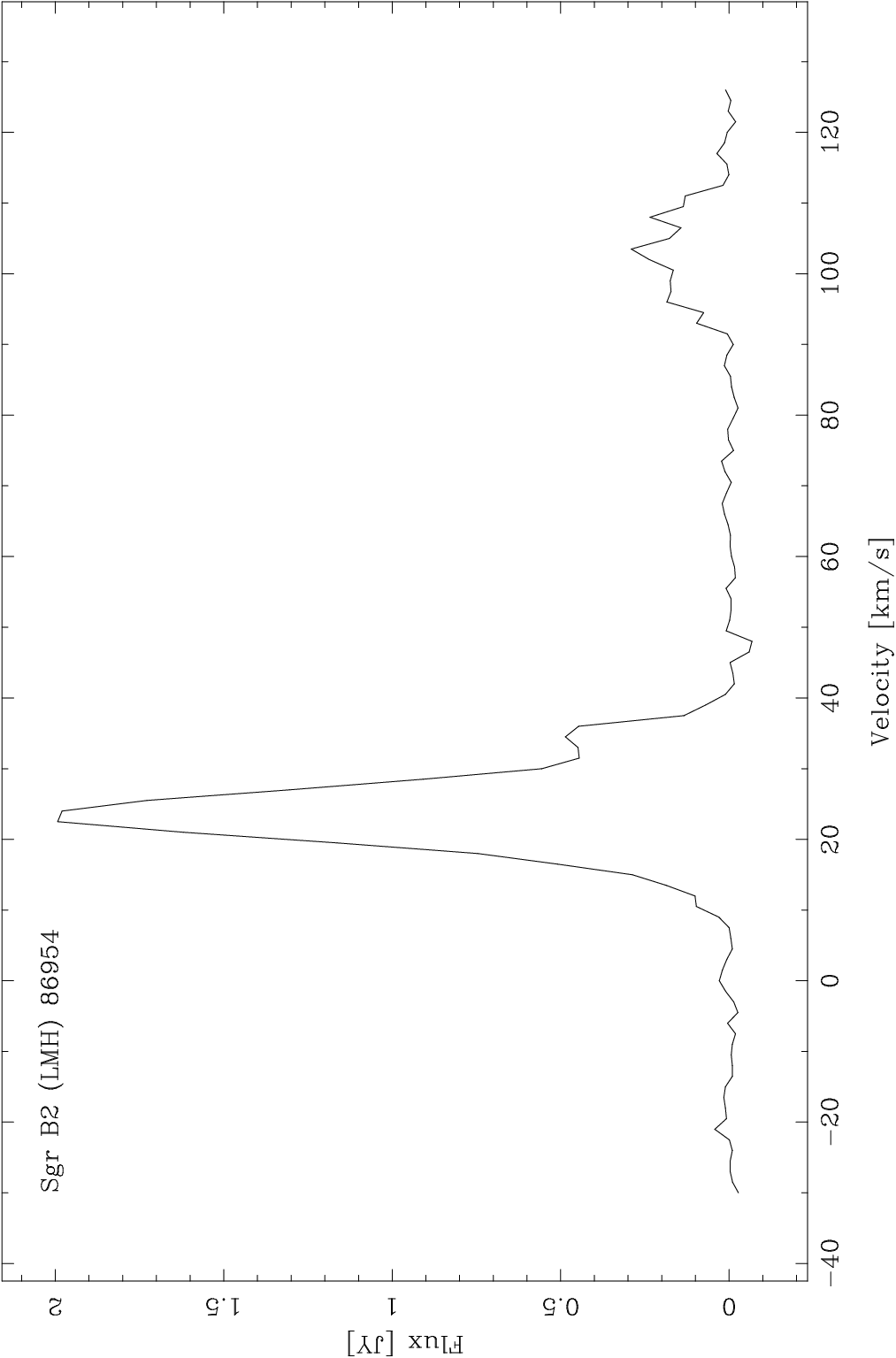}}
\put(14.95,1.1){\vector(0,1){0.3}}
\put(14.6,0.77){\small ? C$_3$H$_7$CN}
\put(11.95,2.45){\vector(0,-1){0.3}}
\put(12.1,2.25){\small ? C$_2$H$_5$OOCH}
\put(11.5,4){\vector(0,1){0.3}}
\put(9.35,4.1){\small ? C$_2$H$_5$OOCH}
\end{picture}
\label{fig2}
\end{figure*}

\begin{figure*}
\setlength{\unitlength}{1cm}
\begin{picture}(16,24)
\put(-1,24.4){\includegraphics{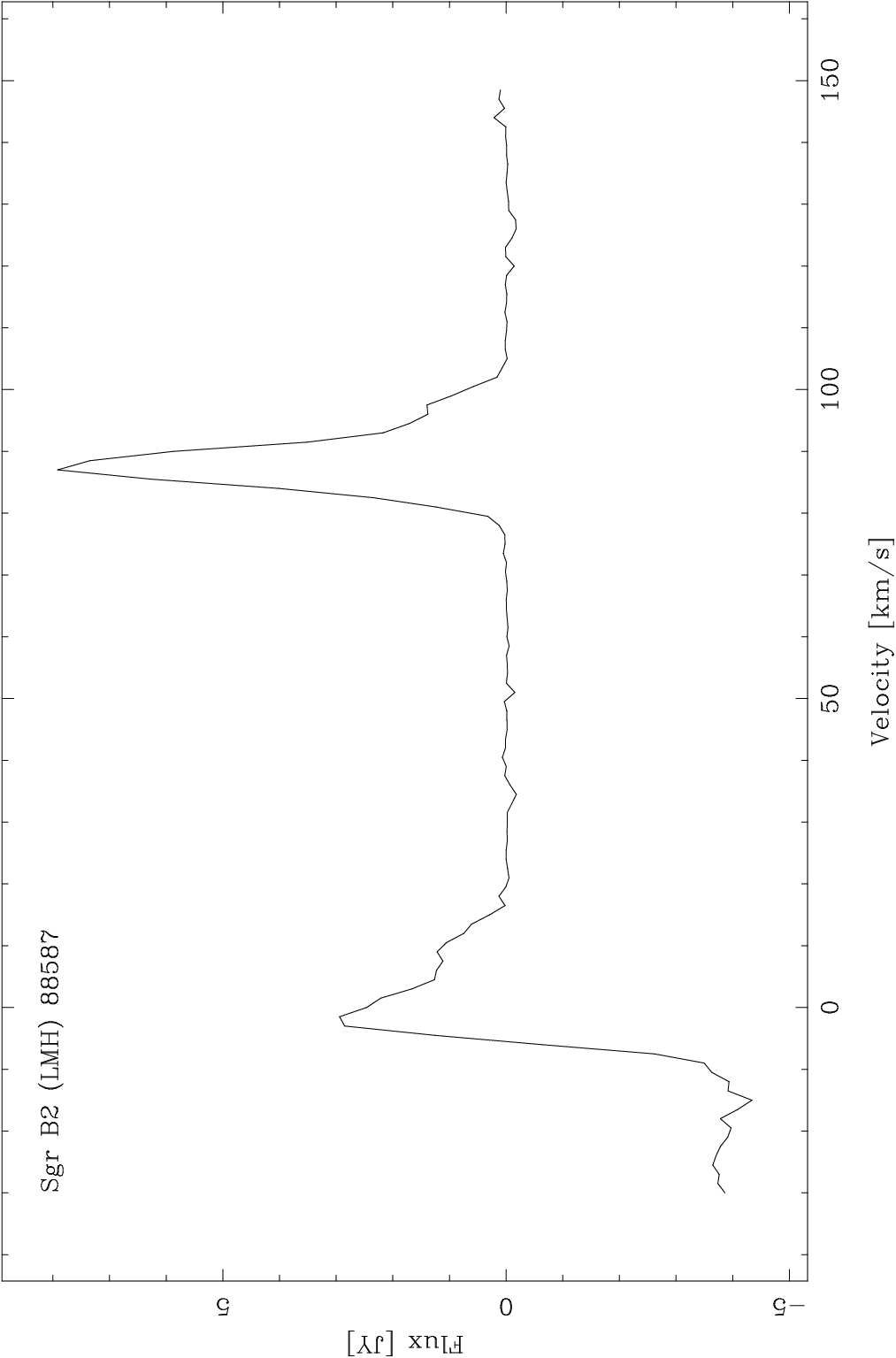}}
\put(5.15,20.4){\vector(0,1){0.4}}
\put(4.75,20.4){\vector(0,1){0.4}}
\put(4.3,20){\small CH$_3$OH}
\put(0.85,19.6){\vector(0,-1){0.4}}
\put(0.1,19.3){\small HCN}
\put(7.8,24.4){\includegraphics{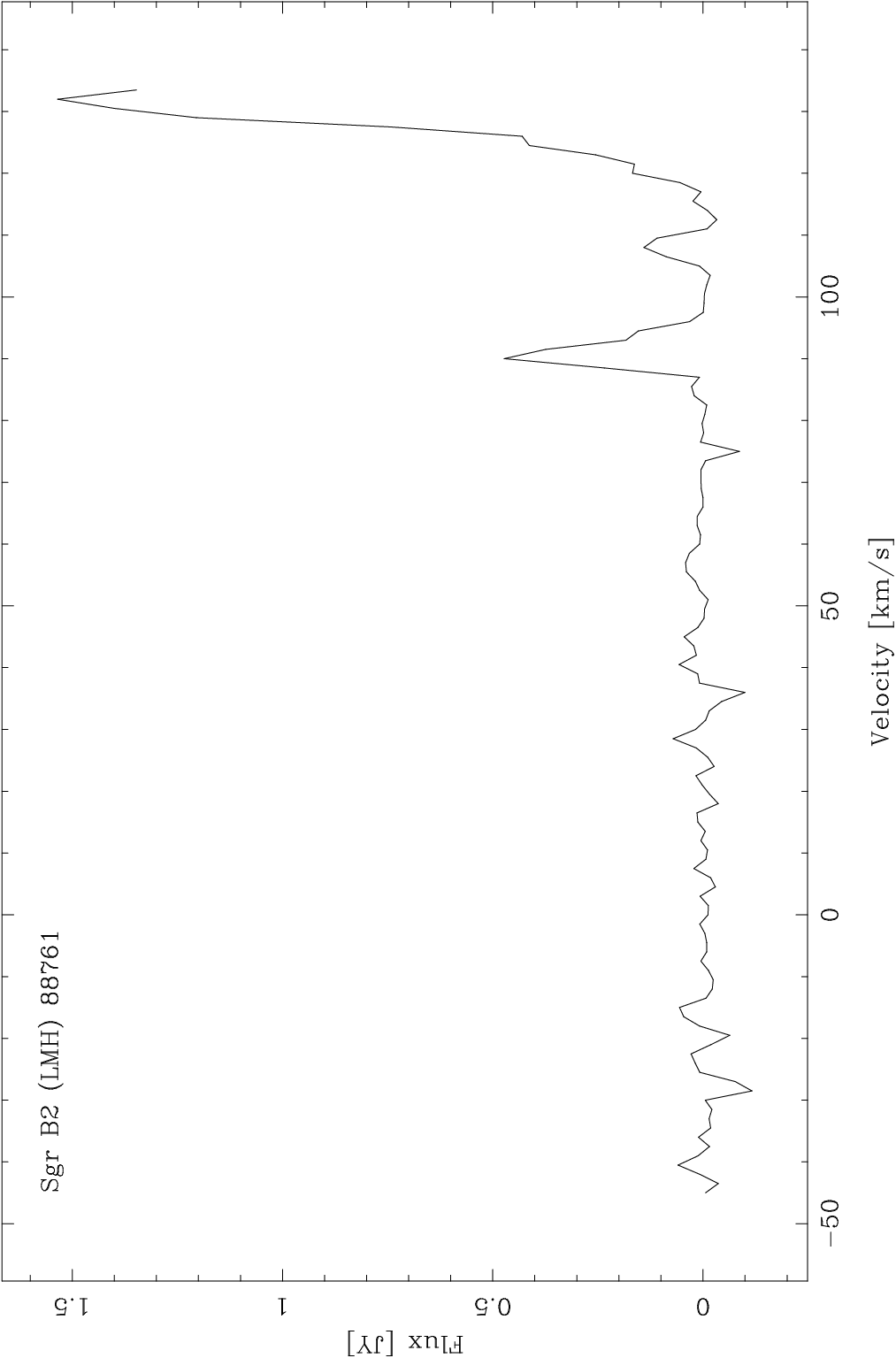}}
\put(15.85,22.1){\vector(0,1){0.4}}
\put(14,22.2){\small CH$_3$CH$_2$CN}
\put(14.95,19.1){\vector(0,1){0.3}}
\put(14.3,18.8){\small ? HOONO$_2$}
\put(14.3,20.9){\vector(0,-1){0.4}}
\put(12.9,20.7){\small U 88.7708}
\put(-1,18.4){\includegraphics{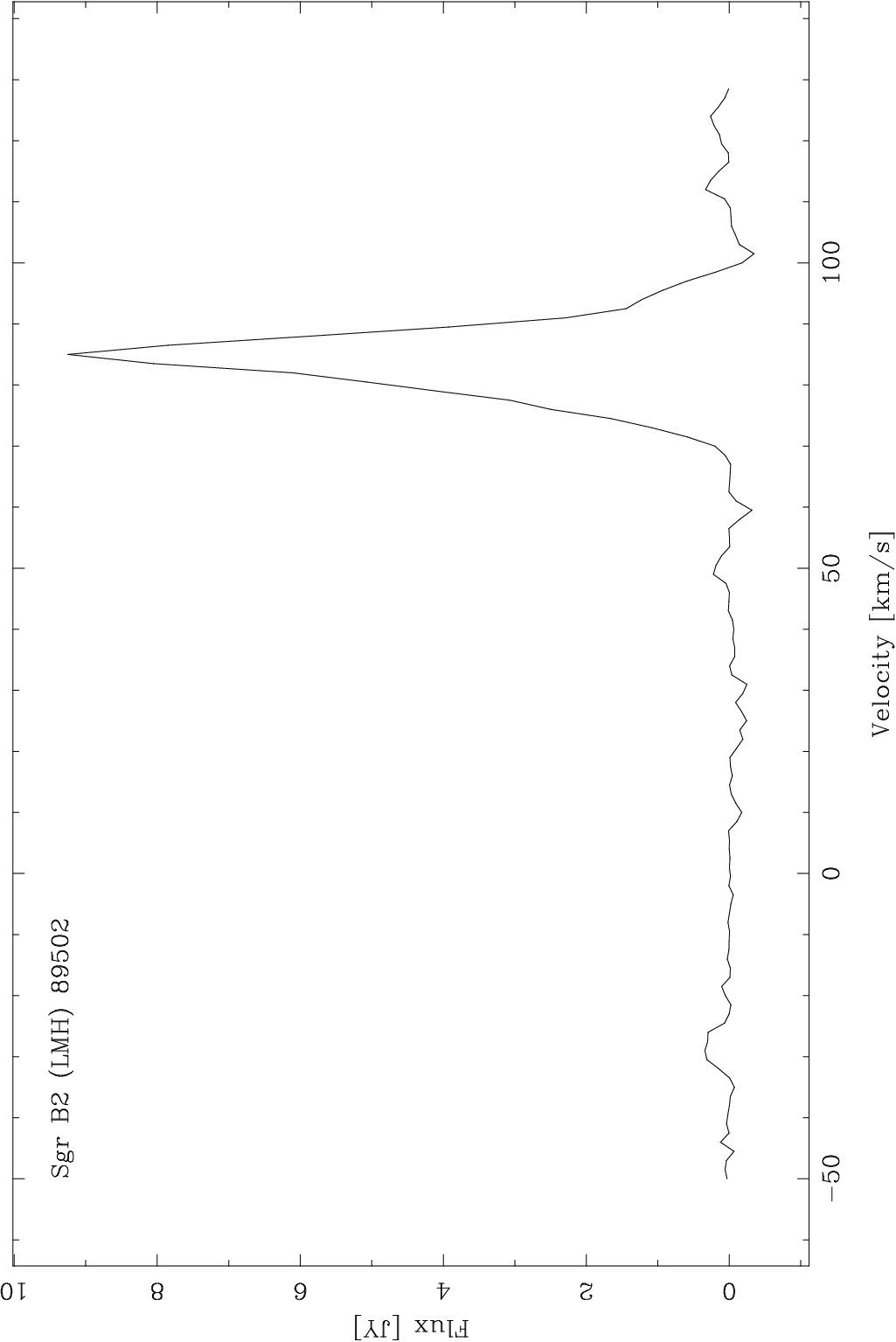}}
\put(6.5,13.7){\vector(0,-1){0.4}}
\put(6.1,13.8){\small ? HOCl}
\put(5.4,13.1){\vector(0,1){0.4}}
\put(5.8,13.1){\vector(0,1){0.4}}
\put(4.85,12.8){\small CH$_3$OH}
\put(4.12,13.7){\vector(0,-1){0.4}}
\put(3.45,13.8){\small ? $^{81}$BrO}
\put(1.2,13.7){\vector(0,-1){0.4}}
\put(0.9,13.8){\small CH$_3$CH$_2$CN}
\put(7.8,18.4){\includegraphics{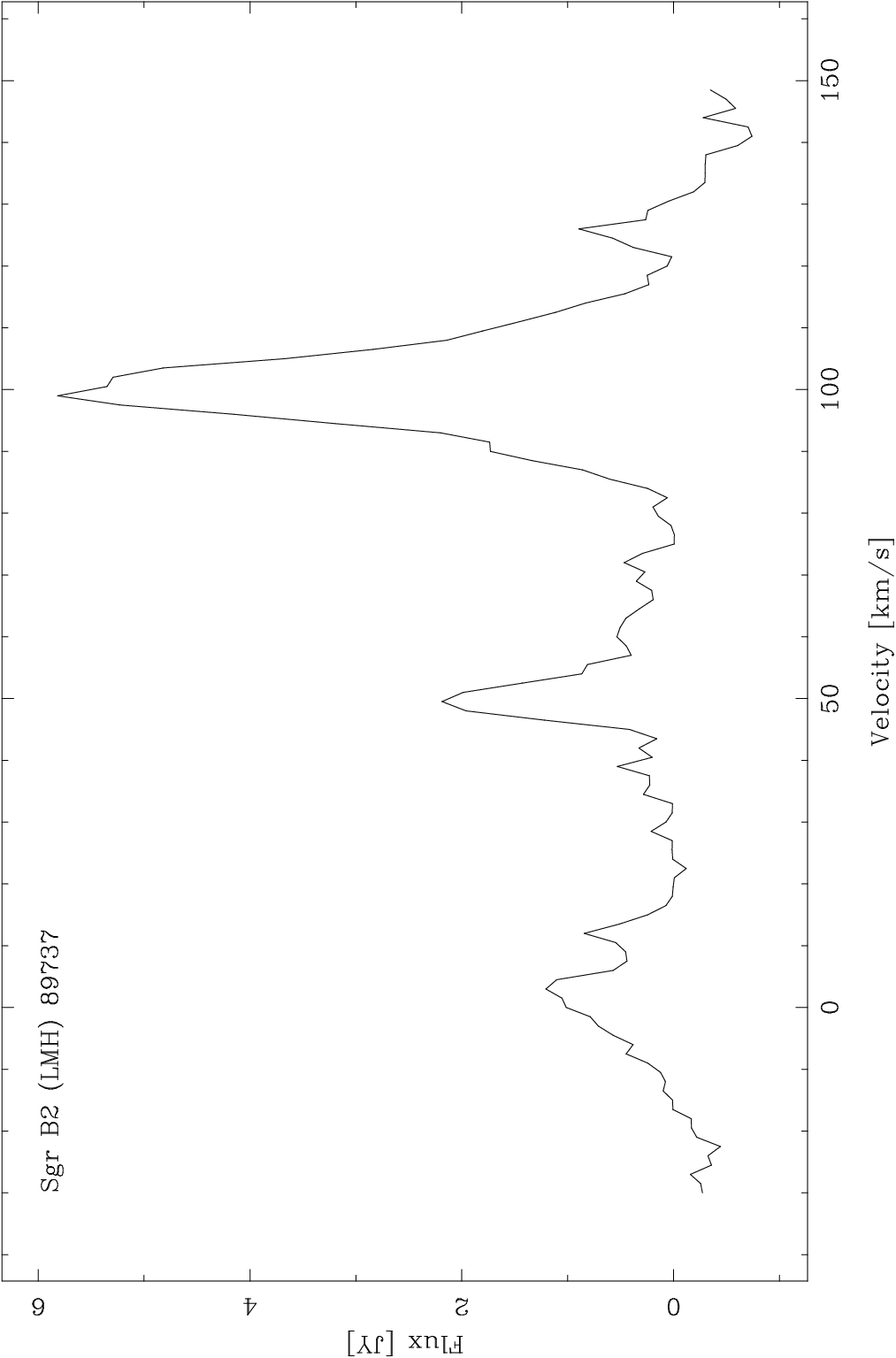}}

\put(-1,12.4){\includegraphics{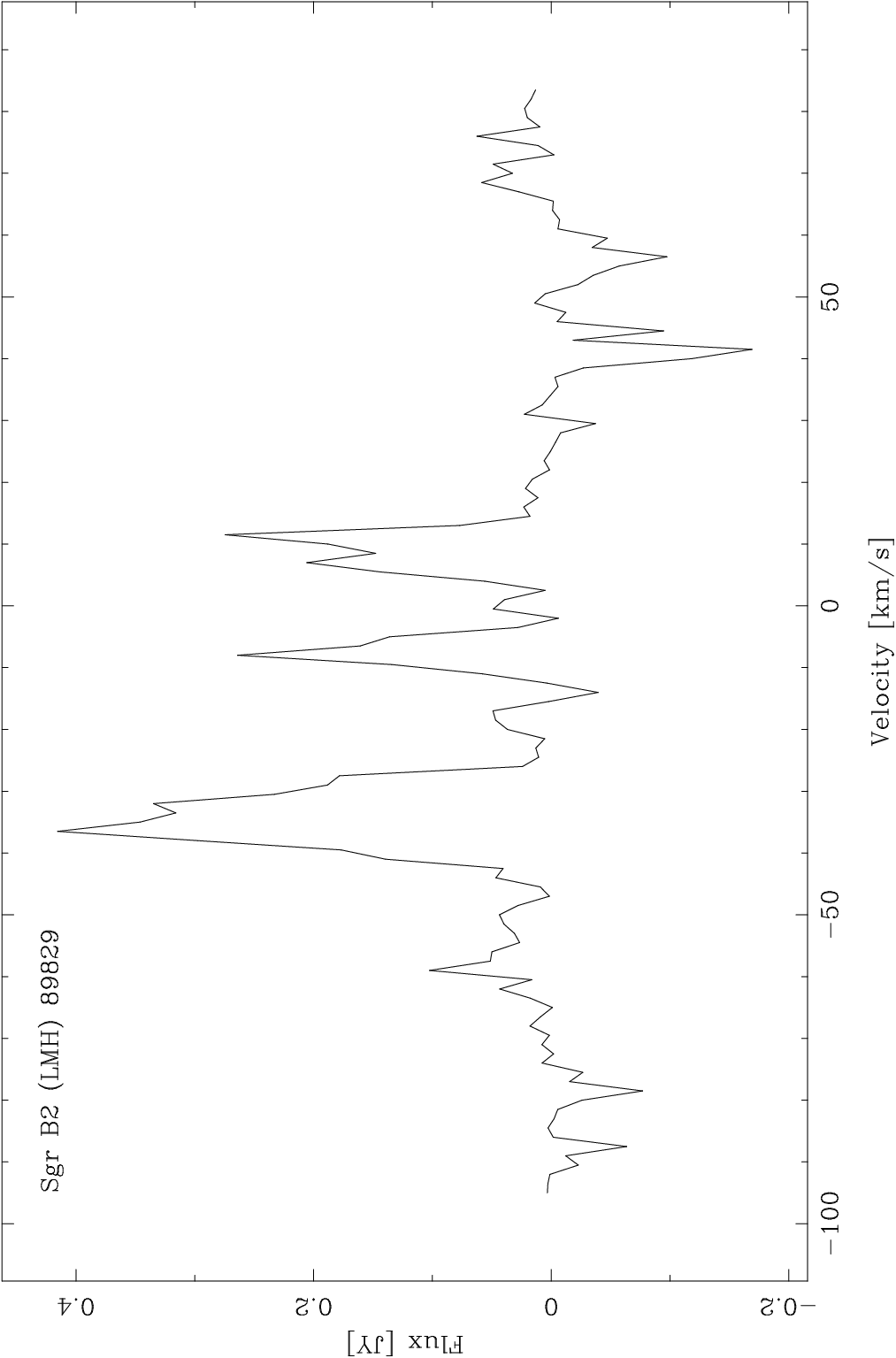}}
\put(2.65,9.8){\vector(0,1){0.3}}
\put(1.3,9.85){\small HCOOH}
\put(7.8,12.4){\includegraphics{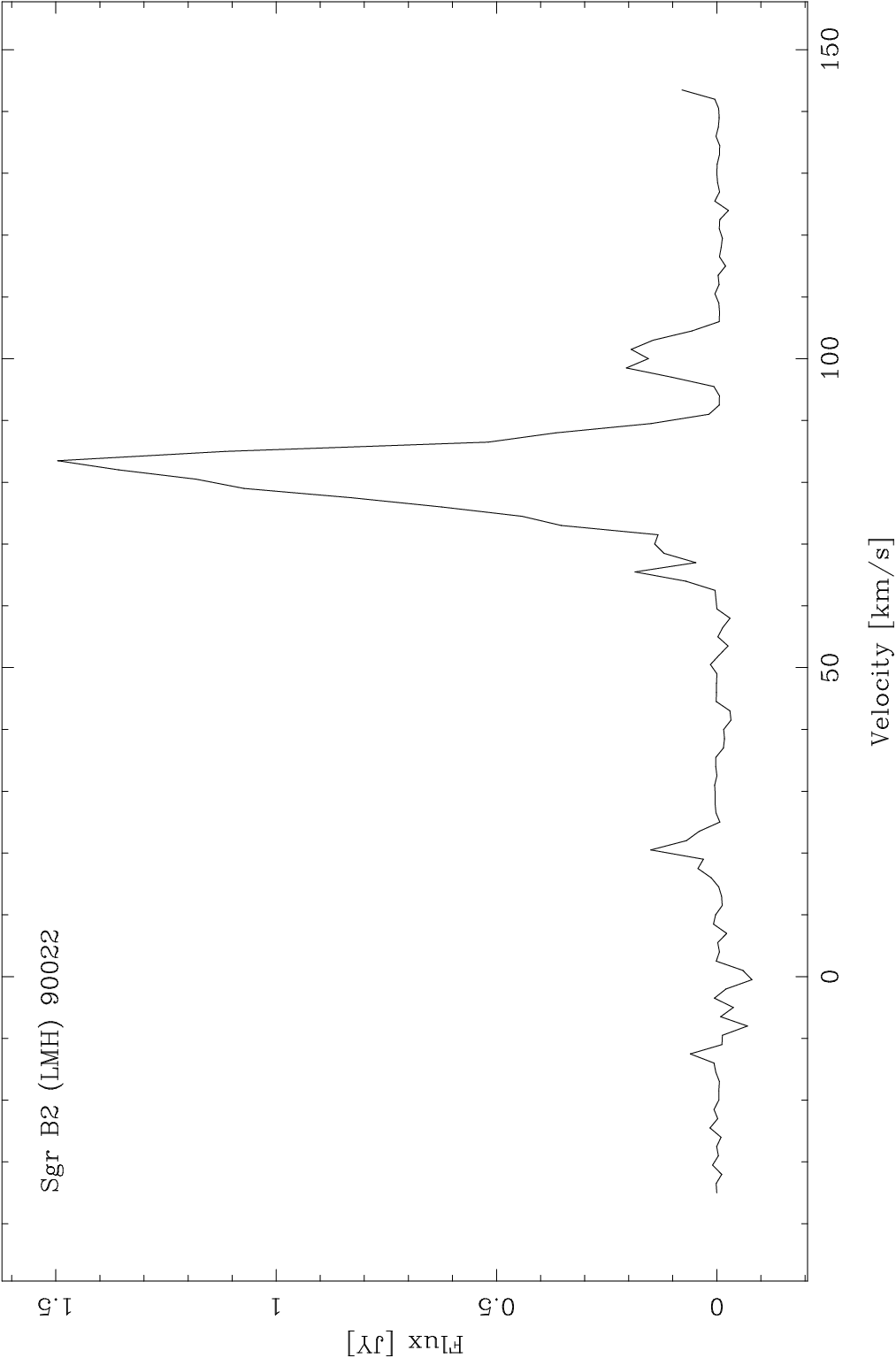}}
\put(13.8,7){\vector(0,1){0.4}}
\put(13.5,7){\vector(0,1){0.4}}
\put(13,6.7){\small U 90.038}
\put(11.3,7){\vector(0,1){0.3}}
\put(10.6,6.7){\small ? C$_3$H$_7$CN}
\put(-1,6.4){\includegraphics{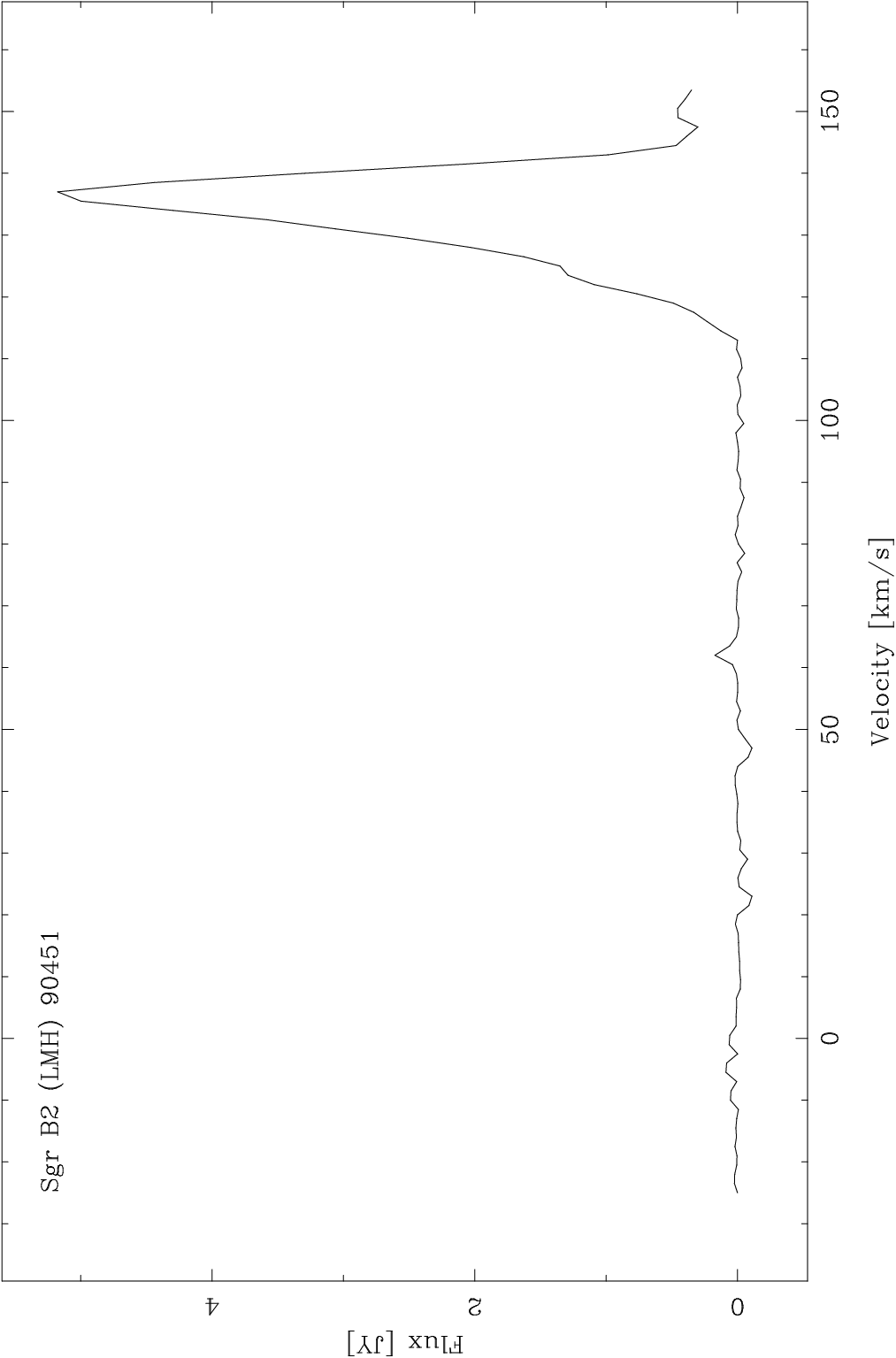}}
\put(6.45,1.2){\vector(0,1){0.3}}
\put(5.5,0.7){\small CH$_3$CH$_2$CN}
\put(7.8,6.4){\includegraphics{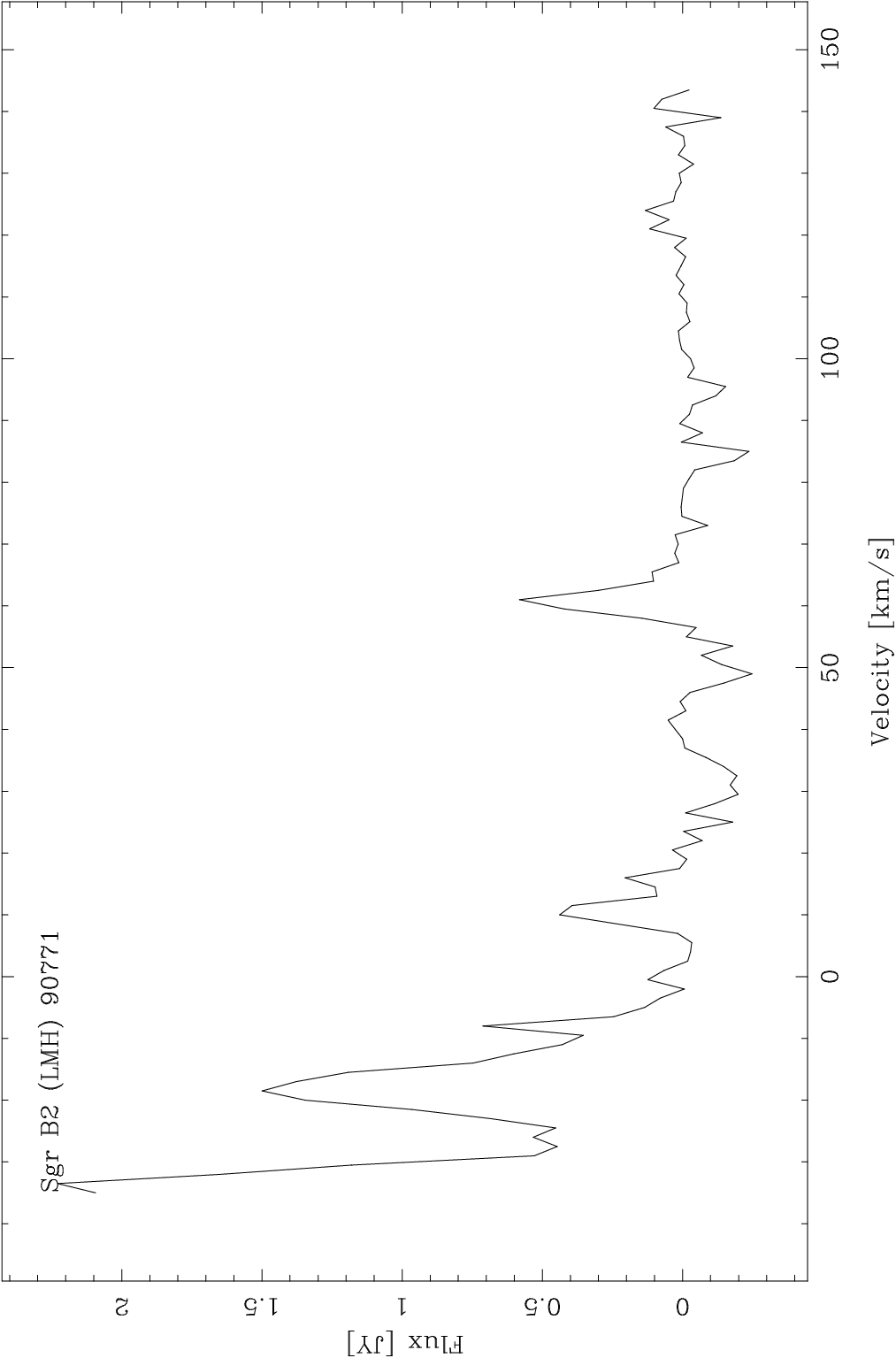}}
\put(12.8,1.5){\vector(0,1){0.3}}
\put(13,1.6){\small ? $^{35}$ClONO$_2$}
\put(11.7,1.5){\vector(0,-1){0.3}}
\put(11.2,1.6){\small ? HNO$_3$}
\put(9.2,4.3){\vector(0,1){0.3}}
\put(9.35,4.35){\small CH$_3$OH}
\end{picture}
\label{fig3}
\end{figure*}



The line features in the data cubes were fitted as Gaussians, using the MIRIAD 
task gaufit, at a pixel near the N continuum position (0.5 arcsec west and 
0.8 arcsec south) where the line emission and absorption was centred. The peak
flux (Jy), width (km/s) and integrated flux (Jy km/s) are quoted in table 3.
To enable comparison with the plots in figure 2, and since the data cubes have 
velocity axes, the fitted line position is given as both a velocity and a frequency. 

The conversion of fitted line velocities to rest frequency requires the
appropriate radial velocity for the emitting region - this is not trivial 
because of the multiple velocity components in Sgr B2.
Lines previously detected in the interstellar medium were identified from
the NIST on-line database of Lovas 
(http://physics.nist.gov/PhysRefData/Micro/Html/ contents.html), and given in 
the last column of table 3. For five known lines where the emission is peaked 
near continuum 
position N, this gives radial velocity 63.5 km/s (with standard deviation 0.7 
km/s). For some of the strong lines in the data cubes, there is a second
component seen  
5 arcsec north of the other lines (0.5 arcsec west and 4.2 arcsec north of the 
N continuum). These lines, labelled d (for double lines) 
in the Identification column of table 3,
were refitted in the data cubes at this more northern position. For the same 
five known lines from the NIST database, these give radial velocity 73.1 km/s
(standard deviation 0.4 km/s) for this position. The rest frequencies quoted in 
table 3 were therefore calculated assuming radial velocity 63.5 km/s for the
position near the N continuum, and 73.1 km/s for the more northerly position,
labelled d in table 3. This spatial and velocity structure is similar to that 
found by Hollis et al. (2003b) in VLA observations of ethyl cyanide at 43.5 GHz,
and other previous observations of Sgr B2 (LMH).

We list, in table 3, 58 emission features and 21 absorption features significant
at the $ 5 \sigma$ level in the 15 chunks of spectrum searched. Assuming 50 MHz
for each band (dropping the edges of the original 64 MHz bands), this gives
around 75 emission features per GHz and 25 absorption features per GHz stronger 
than the median $ 5 \sigma$ level of 110 mJy, after making 
small corrections for the
incompleteness above the 110 mJy level, and the double-counting of the emission
line at 86.745 GHz in the overlap between the 86709 and 87754 bands.  
This is slightly higher than the line density of 61 per GHz in the survey of 
Friedel et al (2004) from BIMA and NRAO 12-m observations of Sgr B2 (N-LMH),
although we have not covered as large a frequency range.
Given the median line width of 6.5 km/s, corresponding to a width of 1.9 MHz,
we find that the spectra have reached a level where there is line confusion,
with around 1/5 of the band covered with lines. The number of emission
features varies with flux density as $n(> S) \propto S^{-0.5}$
and absorption features as $n(> S) \propto S^{-1.0}$.


\begin{table*}
\caption{Emission and absorption lines detected above the 5 $\sigma$ level.
Most are spatially associated with the 64 km/s cloud, and we assumed a cloud velocity of 63.5 km/s
to convert the observed frequency to rest frequency, whereas lines marked `d'
are associated with the 73 km/s cloud and assume a velocity of 73.1 km/s.}
\begin{tabular}{ccrcrccl}
\hline
Band  & Frequency & Velocity & Peak & Width & Integrated &  & Identification \\
Label & (GHz)    & (km/s) & (Jy)   & (km/s) & (Jy km/s) &  &                \\
\hline
      &         &        &         &       &       &   & \\
85477 & 85.4683 &  119.1 &   1.84  &   9.1 & 17.8  &   & ? C$_2$H$_3$CN 85.46862 J \\
      & 85.4727 &  103.7 &  -0.149 &   (5) & -0.79 &   & \\
      & 85.4836 &   65.7 &  -0.149 &   3.6 & -0.57 &   & \\
      & 85.4866 &   55.2 &   0.219 &   3.0 &  0.70 &   & ? O$^{79}$BrO 85.48633 J \\
      & 85.4926 &   34.1 &   0.180 &   6.3 &  1.21 &   & \\
      & 85.4993 &   10.6 &  -0.097 &   3.7 & -0.38 &   & ? HOONO$_2$ 85.49896 J \\
      & 85.5064 &  -14.2 &   0.144 &  10.8 &  1.66 &   & U 85.506 L\\     
      &         &        &         &       &       &   & \\
86267 & 86.2597 &   72.4 &  -0.115 &   (5) & -0.61 &   & \\
      & 86.2631 &   60.5 &   0.34  &   5.1 & 1.84  &   & \\
      & 86.2657 &   51.4 &   1.83  &   5.9 & 11.5  &   & CH$_3$OCHO 86.265826 L \\
      & 86.2686 &   41.4 &   1.43  &   6.3 &  9.6  &   & ? CH$_3$OCHO 86.26865 J \\
      & 86.2738 &   23.4 &   0.59  &   6.9 &  4.3  &   & ? C$_2$H$_5$OOCH 86.27379 J \\
      & 86.2798 &    2.5 &   0.198 &   2.0 &  0.42 &   & \\
      & 86.2850 &  -15.5 &  -0.078 &   9.0 & -0.75 &   & ? NH$_2$CH$_2$CH$_2$OH 86.28495 J \\
      & 86.2918 &  -39.0 &   0.173 &   1.5 & 0.276 &   & ? O$^{37}$ClO 86.29165 J \\
      & 86.3002 &  -58.4 &   0.57  &   (8) &  4.9  & d & ? HOONO$_2$ 86.30031 J \\
      & 86.3005 &  -69.1 &   0.65  &   8.4 &  5.8  &   & ? HOONO$_2$ 86.30031 J \\
      &         &        &         &       &       &   & \\
86374 & 86.3691 &  101.3 &  -0.095 &   7.2 & -0.73 &   & \\
      & 86.3782 &   69.7 &  -0.284 &   6.5 & -1.97 &   & \\
      & 86.3831 &   52.9 &   1.45  &  11.0 &  17.0 &   & ? N$_2$CHO 86.38325 J \\
      & 86.3892 &   31.7 &  -0.61  &   7.3 &  -4.7 &   & ? CH$_2$F$_2$ 86.38924 J \\
      & 86.3986 &    8.6 &   0.73  &   5.1 &  4.0  & d & \\
      & 86.3986 &   -0.8 &   0.62  &   9.3 &  6.2  &   & \\
      &         &        &         &       &       &   & \\
86709 & 86.7097 &  104.7 &   0.118 &   4.2 &  0.53 &   & ? O$^{79}$BrO 86.70978 J \\
      & 86.7453 &   -8.7 &   0.94  &   5.1 &   5.1 & d & CH$_3$CH$_2$CN 86.745317 L \\
      & 86.7453 &  -18.1 &   1.78  &   6.7 &  12.7 &   & CH$_3$CH$_2$CN 86.745317 L \\
      &         &        &         &       &       &   & \\
86754 & 86.7454 &   -9.1 &   0.77  &   5.2 &   4.3 & d & CH$_3$CH$_2$CN 86.745317 L \\
      & 86.7454 &  -18.5 &   1.10  &   9.3 &  10.9 &   & CH$_3$CH$_2$CN 86.745317 L \\
      & 86.7526 &  -43.3 &  -1.34  &  23.8 &   -33 &   & H$^{13}$CO$^+$ 86.754330 L \\
      & 86.7706 & -105.4 &  -0.54  &  13.1 &  -7.5 &   & \\
      & 86.7814 & -142.6 &   0.252 &   3.8 &  1.02 &   & \\
      & 86.7843 & -152.6 &  -0.230 &   2.1 & -0.51 &   & \\
      &         &        &         &       &       &   & \\
86880 & 86.8793 &   86.4 &   0.288 &   9.0 &  2.76 &   & ? O$^{79}$BrO 86.8799 J \\
      & 86.8866 &   61.2 &  -0.36  &   7.0 & -2.71 &   & \\
      & 86.8958 &   29.6 &  -0.40  &   3.6 & -1.52 &   & ? HCOOD 86.89546 J \\
      & 86.9031 &   14.0 &   2.12  &   7.8 &  17.6 & d & CH$_3$OH 86.902947 L \\
      & 86.9029 &    5.2 &   6.6   &   5.2 &  36   &   & CH$_3$OH 86.902947 L \\
      & 86.9150 &  -26.9 &   0.42  &   4.5 &  2.01 & d & \\
      & 86.9152 &  -37.1 &   0.81  &   6.9 &   5.9 &   & \\
      &         &        &         &       &       &   & \\
86954 & 86.9556 &  102.8 &   0.235 &  14.1 &   3.5 &   & ? C$_3$H$_7$CN 86.95546 J \\
      & 86.9784 &   34.1 &   0.99  &   5.3 &   5.6 & d & ? C$_2$H$_5$OOCH 86.97838 J \\  
      & 86.9787 &   23.3 &   2.09  &   9.4 &  20.9 &   & ? C$_2$H$_5$OOCH 86.97838 J \\
      &         &        &         &       &       &   & \\
88587 & 88.5948 &   97.4 &   2.93  &   6.3 &  19.7 & d & CH$_3$OH 88.594809 L \\
      & 88.5946 &   88.3 &   7.5   &   7.5 &    60 &   & CH$_3$OH 88.594809 L \\
      & 88.6191 &    5.8 &   2.78  &  14.9 &    44 &   & \\
      & 88.6215 &   -2.3 &    4.5  &   6.8 &    33 &   & \\
      & 88.6263 &  -18.5 &   -4.2  &  41.3 &  -183 &   & HCN 88.6304157 L \\
      &         &        &         &       &       &   & \\
88761 & 88.7583 &  132.2 &   1.47  &  10.6 &  16.6 &   & CH$_3$CH$_2$CN 88.758419 L \\
      & 88.7655 &  108.1 &   0.151 &   3.3 &  0.53 &   & ? HOONO$_2$ 88.76542 J \\  
      & 88.7706 &   90.7 &   0.45  &   4.5 &  2.17 &   & U 88.7708 L \\
\hline
\end{tabular}
\label{tab3}
\end{table*}

\begin{table*}
\begin{tabular}{ccrcrccl}
\hline
Band  & Frequency & Velocity & Peak & Width & Integrated &  & Identification \\
Label & (GHz)    & (km/s) & (Jy)   & (km/s) & (Jy km/s) &  &                \\
\hline
      &         &        &         &       &       &   & \\
89502 & 89.4946 &  122.3 &   0.188 & (9.6) &  1.92 &   & ? HOCl 89.49483 J \\
      & 89.4974 &  113.1 &   0.227 & (6.4) &  1.55 &   & \\
      & 89.5003 &  103.2 &  -0.158 & (6.4) & -1.08 &   & \\
      & 89.5056 &   95.1 &   2.38  &   5.1 &  12.9 & d & CH$_3$OH 89.505778 L \\
      & 89.5061 &   83.8 &   8.3   &  10.7 &   95  &   & CH$_3$OH 89.505778 L \\
      & 89.5134 &   59.6 &  -0.162 & (6.4) & -1.10 &   & \\
      & 89.5162 &   50.0 &   0.125 &  (8)  &  1.07 &   & ? $^{81}$BrO 89.51593 J \\
      & 89.5232 &   26.6 &  -0.178 &  14.8 & -2.81 &   & \\
      & 89.5397 &  -28.5 &   0.38  &   5.7 &  2.28 &   & CH$_3$CH$_2$CN 89.53945 L \\
      &         &        &         &       &       &   & \\
89737 & 89.7319 &  125.5 &   0.80  &   8.3 &   7.1 &   & \\
      & 89.7395 &  100.2 &   5.4   &  14.7 &   84  &   & \\
      & 89.7546 &   50.1 &   2.06  &   8.3 &  18.2 &   & \\
      & 89.7688 &    2.6 &   0.99  &  16.5 &  17.4 &   & \\
      &         &        &         &       &       &   & \\
89829 & 89.8482 &    9.2 &   0.216 &   7.9 &  1.82 &   & \\
      & 89.8532 &   -7.6 &   0.235 &   4.7 &  1.18 &   & \\
      & 89.8613 &  -34.5 &   0.35  &  (12) &   4.5 &   & HCOOH 89.86148 L \\
      &         &        &         &       &       &   & \\
90022 & 90.0320 &  100.3 &   0.207 &   6.1 &  1.35 &   & \\
      & 90.0375 &   91.8 &   0.33  &  14.6 &   5.1 & d & U 90.038 L \\
      & 90.0378 &   81.1 &   0.97  &   9.8 &  10.1 &   & U 90.038 L \\
      & 90.0560 &   20.8 &   0.138 &   2.9 &  0.43 &   & ? C$_3$H$_7$CN 90.05591 J \\
      &         &        &         &       &       &   & \\
90451 & 90.4535 &  135.1 &   4.7   &  13.0 &    65 &   & CH$_3$CH$_2$CN 90.453354 L \\
      &         &        &         &       &       &   & \\
90771 & 90.7740 &   94.9 &  -0.164 &   2.9 & -0.51 &   & \\
      & 90.7772 &   84.4 &  -0.30  &   2.1 & -0.67 &   & \\
      & 90.7844 &   60.8 &   0.58  &   3.8 &  2.33 &   & ? $^{35}$ClONO$_2$ 90.78419 J \\
      & 90.7877 &   49.8 &  -0.194 &   6.1 & -1.26 &   & \\
      & 90.7935 &   30.8 &  -0.190 &   8.4 & -1.70 &   & ? HNO$_3$ 90.79379 J \\
      & 90.7996 &   10.5 &   0.47  &   3.7 &  1.84 &   & \\
      & 90.8083 &  -17.9 &   1.32  &  12.1 &  17.0 &   & \\ 
      & 90.8132 &  -34.1 &   2.21  &   6.9 &  16.2 &   & CH$_3$OH 90.81239 L \\
\hline
\end{tabular}
\end{table*}


\section{Discussion}

The lines detected (figure \ref{fig2} and table 3) were checked against line frequencies
from the Jet Propulsion Laboratory (JPL, Pickett et al. 1998) and Cologne 
Database for Molecular Spectroscopy (CDMS, 
M\"{u}ller et al 2001) on-line databases. We list in table 3 several lines,
with identifications marked L, where the lines have previously been detected and 
listed in the NIST database of Lovas. We are confident of these identifications.
We also list in table 3, possible identifications based on the JPL on-line 
database, marked with ? and J. These correspond to listed transitions within
0.5 MHz (around 1.7 km/s) of our measured frequencies (using velocity 63.5 km/s for the main
component of Sgr B2 LMH), where the listed frequency uncertainty is less than 
0.5 MHz. Due to the large number of transitions listed in the JPL database,
we have excluded some line identifications with weak transitions 
which would imply implausibly high total column densities (below), and which therefore should have been detected in stronger transitions in other line surveys. We have made 27 tentative assignments in table 3, but many of these
involve molecules not previously detected in the ISM, and cannot be regarded with confidence. Observations of other transitions of these molecules are now necessary to confirm or exclude a detection.  

Using standard assumptions of LTE radiative transfer (eg. Rolfs \& Wilson 2004)
the column density $N_u$ of molecules in the upper level of the transition
is related to the line intensity by
\[ N_u = (8 \pi \nu^2 k/h c^3 A_{ul}) \int T_B dv \]
where $A_{ul}$ is the Einstein coefficient, and $\int T dv$ is the 
integral over velocity of the brightness temperature $T_B$ of the emission line.
The Einstein coefficients $A_{ul}$ were obtained from the Pickett JPL 
or CDMS on-line databases from the tabulated log of intensity $I(T_o)$ at 
reference temperature
$T_o = 300$~K. For some species (CH$_3$OH below) CDMS tabulated values of 
the log of $S_{g} \mu_{g}^2$ are given, and were used to calculate $A_{ul}$.

The total column density $N$ of the molecule is given by
\[ N = (N_u/g_u) Q_T \exp(E_u/kT_{ex}) \]
where $Q_T$ is the partition function at excitation temperature $T_{ex}$, $E_u$
is the energy of the upper level and $g_u$ is the statistical weight of the 
upper level. Values for $g_u$, the lower state energy $E_l$, frequency $\nu$ and
Q as a function of temperature were obtained from the JPL database. 
If several transitions of the same molecule are observed (with 
different upper levels) then this relation can be used, as below 
in a ``rotation diagram'',
to determine both $N$ and $T_{ex}$. 

We did not confidently detect either the glycine or propylene oxide lines. 
We can set $3 \sigma$ upper limits for most of the lines searched. Due to the 
large number of confusing absorption and emission lines, there are several 
observed spectral features that lie close to the lines searched, but not 
close enough to be line identifications. We judge that the absorption lines are 
unlikely to be propylene oxide or glycine identifications, since absorption 
would be due to quite cold dense gas, and that very strong emission lines are
unlikely to be identifications, as they would be inconsistent with the upper
limits given by other transitions not detected.


\begin{table*}
\caption{Column density upper limits for glycine and 
propylene oxide lines searched, at the 3 $\sigma$ level, for those lines 
which were not confused. We label with `conf.' the transitions which were
confused, and give the frequency of the confusing lines. We list flux density and brightness 
temperature limits, molecular line parameters $A_{ul}$, $g_{u}$, $E_{u}/k$, 
and column density limits for the upper level of the transition $N_{u}$ and the total $N$. The calculations have been made assuming that the distribution of the molecules is extended with respect to the ATCA synthezied beam size of $17.0 \times 3.4$ arcsec$^2$. }

\begin{tabular}{lllrcccccc}
\hline
Line Freq.& ID   & Transition              & Limit & Limit & 
$A_{ul}$ & $g_{u}$ & $E_{u}/k$ & $N_{u}$      & $N$  \\
 (GHz)    &      &                         & (mJy)  & (K) & 
($s^{-1}$) &     &   (K)       & (cm$^{-2}$)  & (cm$^{-2}$)  \\  \\ 
\hline
89.51220  & G I  & $13_{6,8} - 12_{6,7}$   &  78 & 0.37 & $ 2.65 \times 10^{-6}$
 & 81  & 41.9 & $ 1.5 \times 10^{13}$ & $ 4.7 \times 10^{15}$ \\
89.53550  & G I  & $13_{6,7} - 12_{6,6}$   &  78 & 0.37 & $ 2.65 \times 10^{-6}$
 & 81  & 41.9 & $ 1.5 \times 10^{13}$ & $ 4.7 \times 10^{15}$ \\
89.83184  & G I  & $14_{1,13} - 13_{1,12}$ &  78 & 0.20 & $ 3.29 \times 10^{-6}$
 & 87  & 34.4 & $ 6.8 \times 10^{12}$ & $ 1.8 \times 10^{15}$ \\
89.87571  & G I  & $13_{5,9} - 12_{5,8}$   &  78 & 0.20 & $ 2.90 \times 10^{-6}$
 & 81  & 38.3 & $ 7.7 \times 10^{12}$ & $ 2.3 \times 10^{15}$ \\
90.04967  & G I  & $15_{0,15} - 14_{0,14}$ &  66 & 0.17 & $ 3.40 \times 10^{-6}$
 &  93 & 35.4 & $ 5.5 \times 10^{12}$ & $ 1.4 \times 10^{15}$ \\
          &      &                         &     &      \\
86.262231 & G II & $12_{6,7} - 11_{6,6}$   &  48 & 0.24 & $ 7.8 \times 10^{-5}$ 
 & 75  & 1044.7 & $ 3.1 \times 10^{11}$ & $ 1.0 \times 10^{14}$ \\
86.283667 & G II & $12_{6,6} - 11_{6,5}$   &  48 & 0.24 & $ 7.8 \times 10^{-5}$
 & 75  & 1044.7 & $ 3.1 \times 10^{11}$ & $ 1.0 \times 10^{14}$ \\
86.716051 & G II & $13_{1,12} - 12_{1,11}$ &  42 & 0.33 & $ 1.0 \times 10^{-4}$
 & 81  & 1037.7 & $ 3.2 \times 10^{11}$ & $ 8.9 \times 10^{13}$ \\
86.960812 & G II & $14_{1,14} - 13_{1,13}$ &  45 & 0.35 & $ 1.1 \times 10^{-4}$
 & 87  & 1038.7 & $ 3.3 \times 10^{11}$ & $ 8.6 \times 10^{13}$ \\
86.967057 & G II & $14_{0,14} - 13_{0,13}$ &  45 & 0.35 & $ 1.1 \times 10^{-4}$
 & 87  & 1038.7 & $ 3.3 \times 10^{11}$ & $ 8.6 \times 10^{13}$ \\
88.792618 & G II & $6_{4,2} - 5_{3,3}$     &  99 & 0.27 & $ 1.9 \times 10^{-6}$
 & 39  & 1018.8 & $ 1.5 \times 10^{13}$ & $ 6.7 \times 10^{15}$ \\
          &      &                         &     &      \\
86.72160  & PO   & $7_{0,7} - 6_{0,6}$     &  42 & 0.33 & $ 3.17 \times 10^{-6}$
 & 15 & 16.8  & $ 1.0 \times 10^{13}$ & $ 1.3 \times 10^{16}$ \\
88.59909  & PO   & $7_{4,4} - 6_{4,3}$     & 105 & 0.50 & $ 2.30 \times 10^{-6}$
 & 15 & 25.9  & $ 2.3 \times 10^{13}$ & $ 3.1 \times 10^{16}$  \\
88.60199  & PO   & $7_{4,3} - 6_{4,2}$     & 105 & 0.50 & $ 2.30 \times 10^{-6}$
 & 15 & 25.9  & $ 2.3 \times 10^{13}$ & $ 3.1 \times 10^{16}$ \\
88.77871  & PO   & $7_{3,4} - 6_{3,3}$     &  99 & 0.27 & $ 2.80 \times 10^{-6}$
 & 15 & 22.0  & $ 1.0 \times 10^{13}$ & $ 1.3 \times 10^{16}$ \\
89.87218  & PO   & $7_{2,5} - 6_{2,4}$     &  78 & 0.20 & $ 3.27 \times 10^{-6}$
 & 15 & 19.4  & $ 6.8 \times 10^{12}$ & $ 8.9 \times 10^{15}$ \\
90.47516  & PO   & $7_{1,6} - 6_{1,5}$     &  93 & 0.24 & $ 3.55 \times 10^{-6}$
 & 15 & 17.9  & $ 7.5 \times 10^{12}$ & $ 9.7 \times 10^{15}$ \\
\hline
86.88593  & G I  & $12_{3,9} - 11_{3,8}$   & \multicolumn{2}{c}{conf. 86.8866 abs.} \\
89.50043  & G I  & $14_{2,13} - 13_{2,12}$ & \multicolumn{2}{c}{conf. 89.5003 abs.} \\
89.75055  & G I  & $13_{4,10} - 12_{4,9}$  & \multicolumn{2}{c}{conf. 89.7546 em.} \\
90.03589  & G I  & $15_{0,15} - 14_{1,14}$ & \multicolumn{2}{c}{conf. 90.0378 em.} \\
90.04311  & G I  & $15_{1,15} - 14_{1,14}$ & \multicolumn{2}{c}{conf. 90.0378 em.} \\
90.05689  & G I  & $15_{0,15} - 14_{0,14}$ & \multicolumn{2}{c}{conf. 90.0560 em.} \\
90.78355  & G I  & $13_{2,11} - 12_{2,10}$ & \multicolumn{2}{c}{conf. 90.7844 em.} \\
          &      &                         &     &      \\
86.379992 & G II & $13_{2,12} - 12_{2,11}$ & \multicolumn{2}{c}{conf. 86.3782 abs.} \\
86.978582 & G II & $12_{5,7} - 11_{5,6}$   & \multicolumn{2}{c}{conf. 86.9787 em.} \\
          &      &                         &     &      \\
85.48422  & PO   & $7_{1,7} - 6_{1,6}$     & \multicolumn{2}{c}{conf. 85.4836 abs.} \\
\hline      
\end{tabular}
\label{tab4}
\end{table*}

\subsection{Upper limits for glycine}

For glycine, conformers I and II, we use the molecular parameters from the JPL
database to calculate the upper limits for the column density of the upper
level of the transition $N_u$, and the total column density $N$.
The results are given in table 4.
Since there are several hyperfine components blended together for 
each line, with different Einstein coefficients $A_{ul}$
and statistical weights $g_u$, we quote $N_u$ based on the main level, with 
$A_{ul}$ and $g_u$ for this level. We include all of the hyperfine
components in the calculation for $N$, by considering $\sum A_{ul} g_u$ 
for all the components. We have assumed excitation temperature 
$T_{ex} = 75$ K, as estimated by Kuan et al. (2003) for glycine I in 
Sgr B2 (LMH), and Q(75) = 29377 for both conformers.

The conformer
II is higher in energy than conformer I by $W_c = 705 $~cm$^{-1}$ (energy 
expressed as $1/\lambda = E/hc$) Lovas et al. (1995) with 
uncertainty 10 \%, or $E/k = 1014 \pm 100$~K. The energy levels of glycine II 
are expressed relative to the lowest energy state of glycine I in the JPL 
database and table 4, but we have considered the two conformers as separate 
species in the total column density calculations. Hence when correcting for the 
Maxwell-Boltzmann distribution of levels we converted the energies to those
relative to the lowest energy state of glycine II for the $\exp(E_u/kT_{ex})$ 
factor, to match the usage in the partition function. The column density limits
obtained for conformer II of glycine are considerably tighter than
those of conformer I due to its larger dipole moment.

The column density limits are averaged over the synthesised beam, and are 
derived from the brightness temperature limits, so they depend on the beam size.
The best upper limit for glycine I from these ATCA observations is 
$ N = 1.4 \times 10^{15}$ cm$^{-2}$ with the beam size of 17.0$\times$3.4 
arcsec$^2$. If the glycine emission is indeed extended over a scale larger 
than the beam, then this non-detection is consistent with the value of  
$ N = 4 \times 10^{14}$ cm$^{-2}$ obtained in Sgr B2 LMH by Kuan et al. (2003)
with the NRAO 12-m.

However, if glycine were confined to the scale of the LMH continuum source 
($<$ 5 arcsec), then the column densities calculated would be beam diluted,
and diluted differently by the different beam sizes. Assuming beam size 
30 arcsec for the NRAO 12-m at 1.3 mm, the beam area ratio is 
$(30 \times 30)/(17 \times 3.4) = 15.6$, so that the ratio of 
ATCA/NRAO 12-m column 
density of $ < 14/4$ or $< 3.5 $ would be $ < 0.22 $ when corrected for
a small scale of glycine emission. 
In other words, if the glycine calculated at a column density of
$ N = 4 \times 10^{14}$ cm$^{-2}$ by Kuan et al. (2003) was distributed
over the scale of a few arcsec (as, for example, methyl formate and acetic acid
are) we would have easily detected it in these ATCA observations.
We therefore conclude that the ATCA 
upper limit provides a strong limit on any small-scale glycine emission, for both of conformers I and II. 

\subsection{Upper limits for propylene oxide}

We have also calculated the upper limits for the column density
of propylene oxide (table 4).
We have assumed excitation temperature 
$T_{ex} = 200$ K, as typical for
Sgr B2 (LMH) as determined by other species (see below section 4.3), 
and Q(200) = 17811 from the the approximate formula
$ Q(T) = (kT/h)^{3/2} (\pi/ABC)^{1/2} $ and the rotational constants
A,B,C from Creswell \& Schwendeman (1977). The energy levels and Einstein 
coefficients (table 4) were calculated with a rigid asymmetric rotor model, 
using the spectroscopic constants from Creswell \& Schwendeman (1977)
and the dipole moment components from Swalen \& Herschbach (1957).


\begin{table*}
\caption{Methanol and ethyl cyanide results.}
\begin{tabular}{llccccc}
\hline
ID and Freq.& Transition & $A_{ul}$ & $g_{u}$ & $E_{u}/k$ & $N_{u}$ & $N_{u}$ \\
 (GHz)      &           & ($s^{-1}$) &   &  (K) & (cm$^{-2}$) & (cm$^{-2}$) \\ 
            &            &           &   &      & 64 km/s & 73 km/s  \\
\hline
CH$_{3}$OH &              &            &     &       &            &         \\ 
86.902947  & $ 7_{2,5} - 6_{3,4} $  & $ 6.85 \times 10^{-7}$ & 15  & 102.7 & 
$ 1.8 \times 10^{16}$ & $ 8.6 \times 10^{15}$ \\
88.594809  & $ 15_{3,13} - 14_{4,10} $ & $ 1.10 \times 10^{-6}$ & 31  & 338.1 & 
$ 1.9 \times 10^{16}$ & $  6.1 \times 10^{15}$ \\
89.505778  & $ 8_{-4,5} - 9_{-3,7} $ & $ 7.65 \times 10^{-7}$  & 17  & 171.4 & 
$ 4.2 \times 10^{16}$ & $ 5.7 \times 10^{15}$ \\
90.81239   & $ 20_{-3,17} - 19_{-2,17}$ & $ 2.81 \times 10^{-6}$  & 41 & 807.9 & 
$ 2.0 \times 10^{15}$   & \\
           &                    &        &   &  &    &    \\ 
CH$_{3}$CH$_{2}$CN &            &        &   &  &    &    \\ 
86.745317  & $ 8_{1,8} - 7_{0,7} $  & $ 3.39 \times 10^{-6}$   & 17  &  16.1 & 
$ 1.2 \times 10^{15}$ & $ 4.7 \times 10^{14}$ \\
88.758419  & $ 27_{3,24} - 27_{2,25} $ & $ 4.89 \times 10^{-6}$ & 55  & 174.0 & 
$ 1.2 \times 10^{15}$ & \\
89.53945   & $ 33_{4,30} - 32_{5,27} $ & $ 1.01 \times 10^{-6}$ & 67  & 259.4 & 
$ 7.6 \times 10^{14}$ & \\
90.453354  & $ 10_{2,8} - 9_{2,7} $  & $ 5.88 \times 10^{-5}$  & 21  &  28.2 & 
$ 3.8 \times 10^{14}$ & \\
\hline      
\end{tabular}
\label{tab5}
\end{table*}

\subsection{Ethyl cyanide and methanol}

Due to the large number of strong transitions of ethyl cyanide 
(CH$_{3}$CH$_{2}$CN) and methanol (CH$_{3}$OH)
in the 3-mm band, we have measured serendipitously 4 transitions
of each molecule. These two molecules are found in the small scale structure
near continuum source N, at around 64 km/s, and also found at
the other component at 73 km/s 5 arcsec to the north. To calculate the
brightness temperature and column density, we assume a size 
$3.5 \times 2.0$ arcsec, based on the median spatial size of line emission 
in our datacubes. The upper level column densities are given in table 5,
along with some of the molecular parameters used. 

Given the detection of several transitions at different upper energy levels, 
we can fit rotation diagrams to derive the excitation temperature and total 
column density. The fits give $T_{ex} = 175$~K and 
$N = 8.4 \times 10^{17}$~cm$^{-2}$ for CH$_{3}$CH$_{2}$CN in the main 64
km/s cloud, $T_{ex} = 190$~K and $N = 1.1 \times 10^{19}$~cm$^{-2}$ 
for CH$_{3}$OH in the 64 km/s cloud and
$T_{ex} = 220$~K and $N = 3.4 \times 10^{18}$~cm$^{-2}$ 
for CH$_{3}$OH in the 73 km/s cloud. 

These excitation temperatures are in good agreement with that expected for 
Sgr B2 LMH, eg. 170 K for CH$_{3}$OH from Pei, Liu \& Snyder (2000). 
The estimates of column density 
are dependent on the assumed source size, but our estimate of total column 
density for CH$_{3}$CH$_{2}$CN in Sgr B2 LMH is about a factor of two higher 
than that obtained by Miao \& Snyder (1997), Liu \& Snyder (1999) 
and Liu, Mehringer \& Snyder (2001) when corrected for the different assumed
sizes. Similarly, our estimate of column density for CH$_{3}$OH is in 
good agreement (30 \% higher) with that of Pei et al. (2000) when corrected 
for source size.

\section{Conclusions}
We have used the ATCA  during 2002 and 2003 to conduct a search for the simplest amino acid, glycine, and the simple chiral molecule propylene oxide, at 3-mm, using the three millimetre capable antennas available at that time. We searched 15 portions of spectrum between 85 and 91 GHz, each of 64 MHz bandwidth. The main results are as follows:
\begin{itemize}
\item We have detected 58 emission features and 21 absorption features in the 15 portions of spectrum searched. This gives a line density of 75 emission and 25 absorption lines per GHz stronger than the 5$\sigma$ level of 110 mJy.
\item We have tentatively assigned 23 of the detected spectral lines to transitions listed in the JPL on-line database (table 3) but as many of these involve molecules not previously detected in the ISM, these assignments cannot be regarded with confidence.  
\item We did not confidently detect either glycine or propylene oxide, but can set 3$\sigma$ upper limits for most transitions searched.
\item Assuming that the glycine emission is extended with respect to the ATCA synthesised beam size of 17.0 x 3.4 arcsec$^2$, our 3$\sigma$ upper limit of $N = 1.4 \times 10^{15}$ cm$^{-2}$ is consistent with that of $N = 4 \times 10^{14}$ cm$^{-2}$ of Kuan et al. (2003) in their reported detection of glycine. However, at the column density reported by Kuan et al. (2003) we show that if glycine were confined to the scale of the LMH continuum source ($<$ 5 arcsec), it would have been easily detected in these ATCA observations. These ATCA observations therefore put a strong upper limit on any small-scale glycine emission in Sgr B2, for both of conformers I and II. 

\end{itemize}

\section*{Acknowledgments}

The Monash authors acknowledge the financial
support provided by the Australia Telescope National Facility for the construction of the laboratory spectrometer, the spectrometer design, construction and technical
services provided by Jonathan G. Crofts, and the collaboration with Takeshi  
Sakaizumi in the laboratory measurements of the Glycine spectrum.
PAJ would like to acknowledge the support provided by  an Australia Telescope National Facility Visiting Fellowship. We also thank the referee for very useful comments that improved
the clarity and presentation of the paper.



\bsp

\label{lastpage}


\end{document}